\documentclass[10pt,aps,pra,twocolumn,superscriptaddress,showpacs]{revtex4-2}
\usepackage{amsmath}
\usepackage{graphicx} % Include figure files
\usepackage{bm}       % bold math
\usepackage{float}
\usepackage{color}
\usepackage[utf8]{inputenc}

\newcommand{\bk}{{\bf k}}
\newcommand{\beq}{\begin{equation}}
\newcommand{\eeq}{\end{equation}}
\newcommand{\bea}{\begin{eqnarray}}
\newcommand{\eea}{\end{eqnarray}}

\begin{document}

\title{Quantum fluctuation induced time of flight correlations of an interacting 
\\ trapped
Bose gas}

\author{Izabella Lovas}
\affiliation{MTA-BME Exotic Quantum Phases ``Momentum'' Research Group\\
Department of Theoretical Physics, Budapest University of Technology and Economics, 1111 Budapest, Budafoki \'ut 8, Hungary}
\author{Bal\'azs D\'ora}
\affiliation{MTA-BME Exotic Quantum Phases ``Momentum'' Research Group\\
Department of Theoretical Physics, Budapest University of Technology and Economics, 1111 Budapest, Budafoki \'ut 8, Hungary}
\author{Eugene Demler}
\affiliation{Physics Departement, Harvard University, Cambridge, Massachusetts 02138, USA}
\author{Gergely Zar\'and}
\affiliation{MTA-BME Exotic Quantum Phases ``Momentum'' Research Group\\
Department of Theoretical Physics, Budapest University of Technology and Economics, 1111 Budapest, Budafoki \'ut 8, Hungary}

%\date{\today}							

\begin{abstract}
We  investigate numerically the momentum correlations in a two dimensional, harmonically trapped interacting 
Bose system at $T=0$ temperature, by using a particle number preserving Bogoliubov approximation. 
Interaction induced quantum fluctuations of the quasi-condensate lead to 
a large anti-correlation dip between particles of wave numbers $\bk$ and $-\bk$  for $|\bk|\sim 1/R_c$, with $R_c$ typical size of the condensate. The anti-correlation dip found is a clear fingerprint of coherent quantum fluctuations of the condensate.   In contrast, for larger wave numbers, $|\bk| \gg 1/R_c$, a weak  positive correlation is found between particles of wave numbers $\bk$ and $-\bk$, in accordance with the Bogoliubov result for homogeneous interacting systems.  
\end{abstract}

\pacs{67.85.-d, 42.50.Lc, 05.30.Jp, 67.85.Hj}

\maketitle

\section{Introduction}

As
 demonstrated first by  Hanbury Brown and Twiss, quantum statistics are efficiently probed through 
 detecting noise correlations.    In their seminal experiments Hanbury Brown and Twiss observed  positive cross-correlations in the shot 
noise of photons emitted by independent light sources~\cite{HBT}. As understood later, 
this photon bunching originates simply from constructive interference between indistinguishable particles, obeying Bose-Einstein 
statistics, and has lately been also demonstrated by interferometry of bosonic atoms ~\cite{HBTboson}. An analogous 
phenomenon is observed for fermions, where the antisymmetry of the wave function results in an antibunching 
behavior~\cite{HBTfermion}. Quantum-statistics related correlations  play an important role in solids, too, 
where  they lead to the emergence of Pauli correlation-hole~\cite{Mahan_book}, 
or  can conspire with interactions to   lead to the emergence of magnetism~\cite{Ashcroft}. 

Measuring Hanbury Brown Twiss-like noise correlations in time of flight (ToF) images  has also been proposed as an efficient tool for detecting correlated states in ultracold atomic systems ~\cite{noisecorr}. Following this suggestion, density correlations in expanding atomic clouds have been used to demonstrate the emergence of ordered phases both in interacting bosonic and fermionic systems~\cite{Schellekens,insitunoise1,insitunoise2,fermionnoise,fermionpairing,
Mottfermion,Mottboson,magneticorder,AForder,HBTBEC,noisereview}, proving that noise detection can also be used to reveal interaction-induced strongly correlated structures.  

Trapped cold atomic systems should provide an ideal test ground to study quantum correlations  in isolated 
bosonic and fermionic systems,  and the influence of interactions on these correlations ~\cite{Dalibard, holograhy,Greiner,lowDgases,Schmiedmayer,BKT}. 
Time of flight  experiments in reduced dimensions~\cite{footnote0}  grant  direct and controlled access 
to the observation of the number $\hat{n}_\mathbf{k}$ of particles with momentum $\hbar \mathbf{k}$ as well as to the 
 correlation function 
 $C(\mathbf{k},\mathbf{k^\prime})\equiv \langle \delta \hat n_\bk \delta \hat n_{\bk '}\rangle$~\cite{experiment,1/k4,boxBEC,2DDalibard,2DDalibard2,Bouchoule2}.

For a very long time~\cite{Altman, ideal, Imambekov, Gangardt},  theoretical predictions regarding the nature of momentum space correlations and ToF correlations in Bose-systems remained  somewhat controversial. 
 %and  it was not entirely clear what precisely one should observe in realistic ToF experiments. 
Two and three dimensional  weakly interacting \emph{homogeneous} systems are quite well-described by a Bogoliubov mean field approximation, where the ground state is found to be a squeezed state generated by the pair creation operators, $\hat{b}^\dagger_\mathbf{k}\hat{b}^\dagger_{-\mathbf{k}}$, 
with $\hat{b}^\dagger_\mathbf{k}$  denoting the creation  operator  of a boson~\cite{Bogoliubov}.  %with wavevector $\bf k$. 
This squeezed structure would imply perfect positive correlations between particles of wave numbers $\mathbf{k}$ and $-\mathbf{k}$ \cite{Altman}. 
However, in a one dimensional Luttinger liquid,  both correlations and anti-correlations  have been predicted~\cite{Altman,Gangardt}, and anti-correlations have also been predicted  
between particles with opposite momenta ~\cite{ideal} 
in harmonically confined noninteracting Bose gases.

\begin{figure}[b!]
\includegraphics[width=0.8\columnwidth]{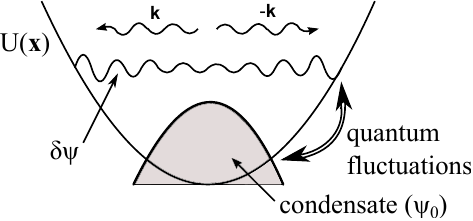}
\caption{Sketch of the origin of quantum fluctuations induced quasiparticle correlations in a trap.  Even at $T=0$, interaction-induced quantum fluctuations of the condensate induce virtual quasiparticle excitations, and amount in fluctuations and correlations, measurable through ToF experiments. The pair structure of excitations induces positive correlation between particles with opposite wave numbers $\mathbf{k}$ and $-\mathbf{k}$.}
\label{fig:sketch} 
\end{figure}

Very recently,  experiments on one-dimensional interacting bosons --- corroborated by detailed theoretical calculations ---
 managed to clarify somewhat  this controversial  
situation ~\cite{Bouchoule2}:  they  confirmed   the predictions of strong \emph{anti-correlations} of Ref.~\cite{Gangardt}
at the momentum scale corresponding to the thermal length, $l_\phi = \rho_{1D} \hbar^2/m k_B T$,  with $\rho_{1D}$ the  density of the one-dimensional gas.  
  
The purpose of the present work is to understand  the role of interaction-induced quantum fluctuations 
of higher dimensional condensates. To be specific, we focus on  $d=2$-dimensional interacting (quasi) condensates, 
where the correlation function $C(\mathbf{k},\mathbf{k^\prime})$  is still directly accessible
experimentally, while a mean field approach is still reliable. Extensions to $d=3$  dimensions are straightforward. 
Focusing  on  interaction-induced quantum fluctuations, we consider the case of $T=0$ temperature only~\cite{footnote1}.

In the presence of interactions,  quantum fluctuations deplete the condensate wave function just as 
thermal fluctuations do in an  ideal gas  (see Fig. \ref{fig:sketch}). 
%G
 Anti-correlations can be interpreted as a sign of conspiracy of particle number conservation and  confinement: they stem from particle number preserving processes, coherently transfering particle pairs between the single mode condensate and the non-condensed fraction of the gas
 (see Sections~\ref{sub:correlation_functions} and ~\ref{sub:simple}).

To capture this physics in a trapped gas, we shall  employ  a particle number preserving Bogoliubov approximation, similar to the one  described in Ref. \cite{Castin}. For sufficiently weak interactions, most of the atoms condense into a single wave function, thereby forming a single-mode condensate $\varphi_0(\mathbf{x})$. Correspondingly,  the bosonic field operator $\hat{\psi}(\mathbf{x})$ can be decomposed as 
\begin{equation}\label{field}
\hat{\psi}(\mathbf{x})=\varphi_0(\mathbf{x})\hat{b}_0+\delta\hat{\psi}(\mathbf{x})\;,
\end{equation}
where $\hat{b}_0$ annihilates a particle from the condensate. 
%G
If the average number of particles in mode $\varphi_0(\mathbf{x})$ greatly exceeds that of non-condensed particles, the operator $\delta\hat{\psi}(\mathbf{x})$, describing quantum fluctuations of the condensate, is small, and  can be accounted for by 
%G 
the particle number conserving mean field approach used here,  
an approach well suited to describing experiments  with a fixed number of particles.

As we shall see, the spatial extension of the condensate ($R_c$)  takes over the role of $l_\phi$
in  one-dimensional condensates~\cite{Bouchoule2}, and determines the region of anti-correlations in momentum space. However, in addition to anti-correlation between small momentum particles with $\bk\approx -\bk'$ and  $|\mathbf{k}|\sim 1/R_c$, a clear \emph{forward correlation} appears for particles of similar momenta,  $\bk\approx \bk'$. Momentum space correlations  thus exhibit  a \emph{p-wave} structure. 
As already explained, these structures are due to interaction induced coherent quantum fluctuations of the condensate, present even at zero temperature.

The expected positive correlations, predicted by Bogoliubov theory, only appear at large wave numbers,  $|\mathbf{k}|\gg 1/R_c$, where  $C(\mathbf{k},-\mathbf{k})$ displays a slowly decaying positive tail of "$d$-wave"-like structure in momentum space. In this large momentum regime, short distance correlations at a scale $\lambda\sim2\pi/|\mathbf{k}|$ are probed, where   correlations can be  well approximated by those  of a 
  homogeneous system. The observation of   Bogoliubov squeezing and the corresponding positive pair correlations  would thus require  investigating the  \emph{tails} of ToF images with high resolution.

The paper is organized as follows:  In Sec.~\ref{sec:methods},  we outline the particle number preserving Bogoliubov approximation following the treatment of Ref. \cite{Castin}, and provide details  on the numerical solution of the corresponding equations (Sec.~\ref{subsec:numerics}). Our results are discussed in Sec.~\ref{sec:result}. Our conclusions are summarized in  Sec.~\ref{sec:cnclusions}.

\section{Methods}
\label{sec:methods}

\subsection{Particle number preserving Bogoliubov approximation}
\label{subsec:bogoliubov}

We  consider a closed, interacting  quasi-two-dimensional Bose gas in a harmonic trap. Such 
quasi-two-dimensional gases can be experimentally realized in highly anisotropic harmonic potentials, where the transverse confinement, $\omega_z$, is much stronger than the trapping frequencies in the remaining two directions \cite{lowDgases}. In this strong vertical confinement limit, the motion of the particles is frozen along the $z$ direction, and the system is described by an effective $d=2$ dimensional Hamiltonian
\bea
H&=& \int \textmd{d}^2\mathbf{x}  \left( \hat{\psi}^{\dagger}(\mathbf{x})\left(-\dfrac{\hbar^2}{2 m}\nabla^2+U(\mathbf{x})\right)\hat{\psi}(\mathbf{x})\right. 
\nonumber
\\
&& \left.+\dfrac{g}{2}\,\hat{\psi}^{\dagger}(\mathbf{x})\hat{\psi}^{\dagger}(\mathbf{x})\hat{\psi}(\mathbf{x})\hat{\psi}(\mathbf{x})\right).
\label{eq:H}
\eea
Here $\hat{\psi}(\mathbf{x})$ denotes the bosonic field operator, and $m$ is the atomic mass. The harmonic potential 
$$U(\mathbf{x})=\dfrac{1}{2}m\omega^2\mathbf{x}^2$$
is responsible for the weak confinement of the atoms in the lateral direction, and the interaction between the atoms is described by a repulsive Dirac-delta potential, $V(\mathbf{x}-\mathbf{x^\prime})= g\,\delta(\mathbf{x}-\mathbf{x^\prime})$ \cite{footnote2}. 
Here the effective interaction $g$ depends sensitively on the vertical confinement, $\omega_z$,  and the three dimensional scattering length $a_{\rm 3D}$~\cite{ZwergerReview}. 
It depends, however,  only logarithmically on the local chemical potential of the Bose gas, and can therefore be replaced by its value at the center 
of the trap for our purposes.

For sufficiently weak interactions, the majority of the atoms condenses into a single wave function, and the system can be analysed by using a Bogoliubov mean field approximation. This approach is justified if the expectation value of the number of non-condensed particles, $\langle\delta\hat{N}\rangle$,
 is only a small fraction of the total particle number $N$,
\bea\label{cond}
\langle\delta\hat{N}\rangle\ll N.
\eea
%G
This condition is necessary for a usual mean field treatment but, in $d=2$ dimensions, considered here, 
 it is not entirely equivalent to the requirement  of weak interactions.
%Condition~\eqref{cond} represents a sufficient but not necessary condition.  
A $d=2$ dimensional  Bose gas can be considered weakly interacting even in the vicinity of the critical (Kosterlitz-Thouless) temperature $T_c$, 
%where $\delta N\sim N$, 
% G: I think this is not true
provided that the dimensionless interaction strength $\tilde{g}$ satisfies \cite{ZwergerReview}
\begin{equation}
\tilde{g}\equiv\dfrac{g\,m}{\hbar^2}\ll 1.
\label{eq:perturbative}
\end{equation}
%G
Standard mean field theory can, however, be applied only in the regime where  where the system size is smaller than the  phase correlation length. 
For typical weakly interacting trapped systems, the latter condition is satisfied only for temperatures $T/T_c\lesssim 0.2$~ \cite{ZwergerReview,PetrovBKT}. 
At slightly larger temperatures, but still below the critical temperature of the Kosterlitz-Thouless phase transition, a so-called quasi-condensate regime appears with large phase fluctuations. Here usual Bogoliubov mean field approach fails, however,  the gradient of the phase still remains small and allows 
a perturbative, generalized Bogoliubov  treatment  \cite{Mora,PetrovBKT}.
At $T\approx 0$, however, condition \eqref{eq:perturbative} is not necessary, and Eq.~\eqref{cond} is satisfied even for slightly 
larger interaction values,  $\tilde{g}\sim1$. 

Below we will concentrate on the regime of true condensate, and will perform calculations at $T=0$ temperature. 
To account for correlations between the condensate and non-condensed particles, 
we shall use a  particle number conserving Bogoliubov approach described in Ref. \cite{Castin}. 
For that purpose, we  decompose the field operator $\hat{\psi}(\mathbf{x})$ according to Eq. \eqref{field}, and separate the single 
mode part $\sim \varphi_0(\mathbf{x})$. The remaining part of the field operator, $\delta\hat{\psi}(\mathbf{x})$, describes  interaction induced 
quantum fluctuations of the condensate (see Fig. \ref{fig:sketch}), and can be chosen to be orthogonal to the wave function $\varphi_0(\mathbf{x})$,
\begin{equation*}
\int \textmd{d}^2\mathbf{x}\, \varphi_0^*(\mathbf{x})\delta\hat{\psi}(\mathbf{x})\equiv 0\,.
\end{equation*}
Next, following Refs.~\cite{Castin,Npreserving}, we introduce a new, particle number preserving field operator 
\begin{equation}\label{lambda}
\hat{\Lambda}(\mathbf{x})\equiv\dfrac{1}{\hat{N}_0^{1/2}}\hat{b}_0^{\dagger}\,\delta\hat{\psi}(\mathbf{x}),
\end{equation}
with ${\hat N}_0\equiv \hat{b}^\dagger_0 \hat{b}_0$ denoting the number of  particles condensed into the single mode part of the 
condensate. The field $\hat{\Lambda}(\mathbf{x})$ satisfies the commutation relations
\begin{align*}
& \left[\hat{\Lambda}(\mathbf{x}),\hat{\Lambda}(\mathbf{x}^{\prime})\right]=0\;, \\
&\left[\hat{\Lambda}(\mathbf{x}),\hat{\Lambda}^{\dagger}(\mathbf{x}^{\prime})\right]=\delta(\mathbf{x}-\mathbf{x}^{\prime})-\varphi_0(\mathbf{x})\varphi_0^*
(\mathbf{x}^{\prime})=\langle\mathbf{x}|\hat{Q}_0|\mathbf{x}^{\prime}\rangle\;,
\end{align*}
with  $\hat{Q}_0\equiv \textmd{Id}-|\varphi_0\rangle\langle\varphi_0|$ denoting the projection onto the subspace orthogonal to $|\varphi_0\rangle$. The operator $\hat{\Lambda}$ transfers one particle from the non-condensed fraction to the condensate, while keeping the total particle number constant. Notice that, in contrast to $\hat{\psi}(\mathbf{x})$, $\hat{\Lambda}(\mathbf{x})$ conserves the particle number, and is therefore more appropriate to describe fluctuations in a closed (microcanonical) trap.

To generate the Gross-Pitaevskii (GP) equation determining the condensate wave function $\varphi_0(\mathbf{x})$, we use the 
ansatz \eqref{field} and approximate the Hamiltonian \eqref{eq:H} by expanding up to second order in the operator 
$\hat{\Lambda}\sim \delta\hat{\psi}$. Particle number conservation is imposed by the exact relations
\bea
N &= &{\hat N}_0 + \delta \hat N\;, 
\nonumber
\\
\delta \hat N &= & \int {\rm d}^2{\bf x}\; \delta \hat{\psi}^\dagger ({\bf x}) \delta \hat{\psi} ({\bf x}) \nonumber  = \int {\rm d}^2{\bf x}\; \hat{\Lambda}^\dagger ({\bf x}) \hat{\Lambda} ({\bf x}),
%+O(\delta\hat{N}/N). 
\nonumber
\eea
which we also assert in course of the expansion. Requiring the disappearance of  terms linear in $\hat{\Lambda}$  yields  the usual Gross-Pitaevskii equation for
$\varphi_0$  
\begin{equation}\label{GP}
\left(-\dfrac{\hbar^2}{2 m}\nabla^2+U(\mathbf{x})\right)\varphi_0(\mathbf{x})+gN|\varphi_0(\mathbf{x})|^2\varphi_0(\mathbf{x})=\mu\varphi_0(\mathbf{x}),
\end{equation}
with  the Lagrange-multiplier $\mu$   introduced to ensure that $\varphi_0$ remain normalized. 
Second order terms in $\hat{\Lambda}$ generate the equation of motion of the field operator,  
\begin{equation*}
i\partial_t
\begin{pmatrix}
\hat{\Lambda}(\mathbf{x}) \\ \hat{\Lambda}^{\dagger}(\mathbf{x})
\end{pmatrix}= \mathcal{L}_{GP}(\mathbf{x})\begin{pmatrix}
\hat{\Lambda}(\mathbf{x}) \\ \hat{\Lambda}^{\dagger}(\mathbf{x})
\end{pmatrix},
\end{equation*}
with the Bogoliubov operator $\mathcal{L}_{GP}$ expressed as  
\begin{align}
\label{LGP}
&\mathcal{L}_{GP}=\begin{pmatrix}
Q_0\left(\mathcal{H}+ g N|\varphi_0|^2\right)Q_0 & g N\,Q_0\,\varphi_0^2\,Q_0^* \\
-g N\,Q_0^*(\varphi_0^*)^2\,Q_0 & \!\!\!\!\! -Q_0^*\left(\mathcal{H}+ g N|\varphi_0|^2\right)Q_0^*
\end{pmatrix},
%\nonumber
\end{align}
and
\begin{equation}
\label{meanH}
\mathcal{H}(\mathbf{x})=-\dfrac{\hbar^2}{2 m}\nabla^2+U(\mathbf{x})-\mu+ g N|\varphi_0(\mathbf{x})|^2
\end{equation}
denoting the mean field single particle Hamiltonian. The Lagrange-multiplier $\mu$ appears here as a chemical potential, 
expressing that the condensate serves as a particle reservoir for the non-condensed fraction of the gas.

The eigenvalues and eigenvectors of the non-Hermitian operator $\mathcal{L}_{GP}$ 
determine the excitation modes of the condensate. The Bogoliubov operator $\mathcal{L}_{GP}$ has a pair of zero-modes \cite{LGP, Castin}
$$ (\varphi_0(\mathbf{x}),0),\quad (0, \varphi_0^*(\mathbf{x})) $$ 
corresponding to -- physically meaningless -- global  phase  rotations of the condensate. All other, nonzero eigenvalues of $\mathcal{L}_{GP}$ come in pairs, $\pm\varepsilon_s$, and correspond to quasiparticle excitations. By denoting the eigenvector of positive eigenvalue $\varepsilon_s>0$ ($s=1,2,...$) by $(u_s(\mathbf{x}), v_s(\mathbf{x}))$, we find that $(v_s^*(\mathbf{x}),u_s^*(\mathbf{x}))$ is also an eigenvector of eigenvalue $\varepsilon_{-s}=-\varepsilon_s$. The positive eigenvectors of $s,s^\prime>0$ satisfy the orthogonality condition
$$\displaystyle\int \textmd{d}^2\mathbf{x} \left(u_s^*(\mathbf{x})u_{s^\prime}(\mathbf{x})-v_s^*(\mathbf{x})v_{s^\prime}(\mathbf{x})\right)=\delta_{s,s^\prime}.$$
Moreover, together with the condensate wave function they form a complete basis, expressed by the relation
\begin{align}\label{complete}
&\sum_{\epsilon_s>0}\left(u_s(\mathbf{x})u_s^*(\mathbf{x^\prime})-v_s^*(\mathbf{x})v_s(\mathbf{x^\prime})\right) \nonumber \\
&\quad\quad+\varphi_0(\mathbf{x})\varphi_0^*(\mathbf{x^\prime})=\delta(\mathbf{x}-\mathbf{x^\prime}).
\end{align}
These eigenfunctions of $\mathcal{L}_{GP}$ can then be naturally used to expand the field operator $\hat{\Lambda}(\mathbf{x})$  as
\begin{equation}\label{expans}
\hat{\Lambda}(\mathbf{x})=\sum_{\varepsilon_s>0}\left[\hat{b}_s\, u_s(\mathbf{x})+\hat{b}_s^{\dagger}\, v_s^*(\mathbf{x})\right],
\end{equation}
where the $\hat{b}_s$'s satisfy bosonic commutation relations and annihilate quasiparticles of (positive) energy $\varepsilon_s$. In terms of these quasiparticle excitations, within the Bogoliubov approximation, the Hamiltonian takes on  a simple diagonal form
\begin{equation*}
H=E_0+\displaystyle \sum_{\varepsilon_s>0}\varepsilon_s\hat{b}_s^\dagger\hat{b}_s\,.
\end{equation*}
%G
The ground state of the system is thus simply the vacuum state of the annihilation operators $\hat{b}_s$. We remark that 
the ground state energy, $E_0$, incorporates interaction dependent negative corrections to the  Gross-Pitaevski mean field energy, 
resulting from   the  quantum depletion of the condensate.

Let us now turn to the computation of  the expectation value $\langle \hat{n}_\bk \rangle$ and the correlation function $\langle \hat{n}_\bk \hat{n}_{\bk'} \rangle$. The particle number operator $\hat{n}_{\mathbf{k}}$ corresponding to wave number $\mathbf{k}$ is defined as
$$\hat{n}_{\mathbf{k}}=\hat{\psi}_{\mathbf{k}}^\dagger\hat{\psi}_{\mathbf{k}},$$
where $\hat{\psi}_{\mathbf{k}}$ is the Fourier-transform of the field operator,
$$\hat{\psi}_{\mathbf{k}}=\int\textmd{d}^2\mathbf{x}\,e^{-i\mathbf{k}\mathbf{x}}\hat{\psi}(\mathbf{x}).$$
In order to calculate the expectation value and correlation function of the operator $\hat n_\bk$, we use Eqs. \eqref{field} and \eqref{lambda} to express $\hat{n}_{\mathbf{k}}$ in terms of the operator $\hat{\Lambda}$, to find
\begin{eqnarray}
\hat{n}_{\mathbf{k}}=& N|\varphi_0(\mathbf{k})|^2-|\varphi_0(\mathbf{k})|^2\,\delta\hat{N} +\sqrt{N}\varphi_0^*(\mathbf{k})\,\hat{\Lambda}_{\mathbf{k}} 
\label{n_k}
\\
&+\sqrt{N}
\varphi_0(\mathbf{k})\,\hat{\Lambda}_{\mathbf{k}}^{\dagger}+\hat{\Lambda}_{\mathbf{k}}^{\dagger}
\hat{\Lambda}_{\mathbf{k}}
+{\cal O}({\delta \hat N}^{3/2}N^{-1/2}),
\nonumber
\end{eqnarray}
with $\hat{\Lambda}_{\mathbf{k}}$ denoting the Fourier transform of $\hat{\Lambda}$,
\begin{equation*}
\hat{\Lambda}_{\mathbf{k}}=\displaystyle\sum_{\varepsilon_s>0}\left[\hat{b}_s\, u_s(\mathbf{k})+\hat{b}_s^{\dagger}\, v_s^*(-\mathbf{k})\right].
\end{equation*}
%G
Notice  that the second term in Eq.~\eqref{n_k} does not appear in  the usual Bogoliubov approach. It is a direct consequence of the particle number conserving method, and leads to corrections in the 
expressions of the correlation functions. This term may be contrasted to the third and fourth terms, which are also related to particle number 
conserving processes but appear already within the usual  Bololiubov approach; 
these describe  the annihilation (creation) of a particle in the cloud of quantum fluctuations, 
while adding (removing) a particle to the condensate (from the condensate).   

Notice that the usual and heuristic identification,  $\hat n_\bk \leftrightarrow \hat{\Lambda}_{\mathbf{k}}^{\dagger}
\hat{\Lambda}_{\mathbf{k}}  $ is not appropriate for a trapped microcanonical condensate, where correlations 
between the single mode part of the condensate and $\delta \hat\psi ({\bf x})$ cannot be neglected. For a homogeneous condensate, 
however, $\varphi^{\rm hom}_0(\bk\ne0)\equiv 0$, and  Eq.~\eqref{n_k} reduces to the simple relation, 
$\hat n^{\rm hom}_{\bk\ne0} =\hat{\Lambda}_{\mathbf{k}}^{\dagger}
\hat{\Lambda}_{\mathbf{k}}  $.

The ground state expectation value of $\hat{n}_{\mathbf{k}}$ is thus  given in terms of eigenfunctions $(u_s(\mathbf{x}), v_s(\mathbf{x}))$ as
\begin{align}\label{nk}
\langle n_{\mathbf{k}}\rangle= & N|\varphi_0(\mathbf{k})|^2+\sum_{\varepsilon_s>0} |v_s(-\mathbf{k})|^2 \nonumber\\
& -|\varphi_0(\mathbf{k})|^2\sum_{\varepsilon_s>0}\int d^2\mathbf{x}\,|v_s(\mathbf{x})|^2.
\end{align}
Here the first term is simply the Gross-Pitaevskii result, describing a situation when all particles belong to the single-mode condensate. The sum $\sum_s |v_s(-\mathbf{k})|^2$ takes into account the contribution of the non-condensed fraction of the gas, while the  last term originates from the depletion of the condensate due to particle number conservation.
Similarly, the correlation function of $\hat{n}_{\mathbf{k}}$ and $\hat{n}_{\mathbf{k^\prime}}$ operators can be expressed as
\begin{widetext}
\begin{align*} 
C(\mathbf{k},\mathbf{k}^{\prime})=&\langle\hat{\psi}^{\dagger}_{\mathbf{k}}\hat{\psi}_{\mathbf{k}}
\hat{\psi}^{\dagger}_{\mathbf{k}^{\prime}}\hat{\psi}_{\mathbf{k}^{\prime}}\rangle
-\langle\hat{\psi}^{\dagger}_{\mathbf{k}}\hat{\psi}_{\mathbf{k}}\rangle\langle
\hat{\psi}^{\dagger}_{\mathbf{k}^{\prime}}\hat{\psi}_{\mathbf{k}^{\prime}}\rangle= \nonumber\\
& N\,\sum_s\left(\varphi_0^*(\mathbf{k})\,u_s(\mathbf{k})+\varphi_0(\mathbf{k})\,v_s(-\mathbf{k})
\right)\left(\varphi_0(\mathbf{k}^{\prime})\,u_s^{*}(\mathbf{k}^{\prime})+\varphi_0^*(\mathbf{k}^{\prime})\,v_s^{*}(-\mathbf{k}^{\prime})\right) \nonumber\\
& +\sum_{s_1 ,\,s_2,\,s_3,\,s_4}\left(\delta_{s_1,\,s_4}\delta_{s_2,\,s_3}+\delta_{s_1,\,s_3}\delta_{s_2,\,s_4}\right)\left(v_{s_1}(-\mathbf{k})u_{s_2}(\mathbf{k})-|\varphi_0(\mathbf{k})|^2\!\int\! \textmd{d}^2 \mathbf{x}\,v_{s_1}(\mathbf{x})u_{s_2}(\mathbf{x})\right)\cdot \nonumber\\
& \left(v_{s_4}^*(-\mathbf{k}^{\prime})u_{s_3}^*(\mathbf{k}^{\prime})-|\varphi_0(\mathbf{k}^{\prime})|^2\!\int\! \textmd{d}^2 \mathbf{x}\,v_{s_4}^*(\mathbf{x})u_{s_3}^*(\mathbf{x})\right).
%\label{eq:correlator}
\end{align*}
\end{widetext}
This equation can be rewritten in a form more convenient for numerical calculations, using the completeness relation Eq. ~\eqref{complete}. Expressing $\sum_s u_s(\mathbf{k})u_s^*(\mathbf{k^\prime})$ from the Fourier transform of Eq. ~\eqref{complete} allows us to separate the singular, $\sim\delta(\mathbf{k}-\mathbf{k^\prime})$ terms appearing in the diagonal correlation function $C(\mathbf{k},\mathbf{k})$. As a result, the correlation function can be written as a sum of three contributions\begin{align}\label{corr}
&C(\mathbf{k},\mathbf{k}^{\prime})=(2\pi)^2\delta(\mathbf{k}-\mathbf{k^\prime})\langle\hat{n}_{\mathbf{k}}\rangle \nonumber\\
&\quad\quad+ C^{(1)}(\mathbf{k},\mathbf{k}^{\prime})+C^{(2)}(\mathbf{k},\mathbf{k}^{\prime}),
\end{align}
with $\langle\hat{n}_{\mathbf{k}}\rangle$ given by Eq. ~\eqref{nk}, and
\begin{widetext}
\begin{subequations}
\begin{align}
&
%C^{(1)}
C^{(1)}(\mathbf{k},\mathbf{k}^{\prime})\equiv N\,\sum_s[\varphi_0^*(\mathbf{k})\varphi_0^*(\mathbf{k}^{\prime})\,u_s(\mathbf{k})\,v_s^{*}(-\mathbf{k}^{\prime})+\varphi_0(\mathbf{k})\varphi_0(\mathbf{k}^{\prime})\,v_s(-\mathbf{k})\,u_s^{*}(\mathbf{k}^{\prime})+\varphi_0(\mathbf{k})\varphi_0^*(\mathbf{k}^{\prime})\,v_s(-\mathbf{k})\,v_s^{*}(-\mathbf{k}^{\prime}) \nonumber \\
&\hspace{5cm}+\varphi_0^*(\mathbf{k})\varphi_0(\mathbf{k}^{\prime})\,v_s^*(-\mathbf{k})\,v_s(\mathbf{-k}^{\prime})]-N |\varphi_0(\mathbf{k})|^2|\varphi_0(\mathbf{k^\prime})|^2, \label{corrterm1} \\
&%C^{(2)}
C^{(2)}(\mathbf{k},\mathbf{k}^{\prime})\equiv\sum_{s_1 ,\,s_2}\left(v_{s_1}(-\mathbf{k})u_{s_2}(\mathbf{k})-|\varphi_0(\mathbf{k})|^2\!\int\! \textmd{d}^2 \mathbf{x}\,v_{s_1}(\mathbf{x})u_{s_2}(\mathbf{x})\right)\left(v_{s_2}^*(-\mathbf{k}^{\prime})u_{s_1}^*(\mathbf{k}^{\prime})-|\varphi_0(\mathbf{k}^{\prime})|^2\!\int\! \textmd{d}^2 \mathbf{x}\,v_{s_2}^*(\mathbf{x})u_{s_1}^*(\mathbf{x})\right) \nonumber \\
&\quad+\sum_{s_1 ,\,s_2}\left(v_{s_1}(-\mathbf{k})v_{s_2}^*(-\mathbf{k})-|\varphi_0(\mathbf{k})|^2\!\int\! \textmd{d}^2 \mathbf{x}\,v_{s_1}(\mathbf{x})v_{s_2}^*(\mathbf{x})\right)\left(v_{s_1}^*(-\mathbf{k}^{\prime})v_{s_2}(-\mathbf{k}^{\prime})-|\varphi_0(\mathbf{k}^{\prime})|^2\!\int\! \textmd{d}^2 \mathbf{x}\,v_{s_1}^*(\mathbf{x})v_{s_2}(\mathbf{x}) \right)\nonumber \\
&\quad-\varphi_0(\mathbf{k})\varphi_0^*(\mathbf{k^\prime})\sum_s v_s(-\mathbf{k})v_s^{*}(-\mathbf{k}^{\prime})-|\varphi_0(\mathbf{k})|^2\sum_s|v_s(-\mathbf{k^\prime})|^2-|\varphi_0(\mathbf{k^\prime})|^2\sum_s|v_s(-\mathbf{k})|^2\nonumber \\
&\quad+|\varphi_0(\mathbf{k})|^2 |\varphi_0(\mathbf{k^\prime})|^2\sum_s\int\! \textmd{d}^2 \mathbf{x}\,|v_{s}(\mathbf{x})|^2.
\label{corrterm2}
\end{align}
\end{subequations}
\end{widetext}
Here, besides Eq. ~\eqref{complete}, we have used that the eigenfunctions $u_s$ and $v_s^*$ are orthogonal to the condensate wave function $\varphi_0$.

%G
The first term in Eq. ~\eqref{corr} denotes the  shot noise. The first correction, $C^{(1)}(\mathbf{k},\mathbf{k}^{\prime})$, is proportional to the total particle number $N$, and includes terms of second order in fluctuations, ${\cal O}(|\delta \psi|^2)$, describing correlations between the single mode condensate and the non-condensed part of the wave function \cite{footnote5}. The second  correction, $C^{(2)}(\mathbf{k},\mathbf{k}^{\prime})$, is of fourth order in fluctuations, ${\cal O}(|\delta \psi|^4)$, and takes into account correlations inside the non-condensed cloud and subleading corrections to the condensate - quasiparticle correlations contained in $C^{(1)}$. These latter are generated by the second term in Eq.~\eqref{n_k}, and account for the depletion 
of the single mode condensate. % \cite{footnote7}.  \bibitem{footnote7}
The "cylindrically symmetrical" terms in Eq.~\eqref{corrterm2}, proportional to $|\varphi_0(\mathbf{k})|^2$ (or $|\varphi_0(\mathbf{k^\prime})|^2$), stem from   correlations between the condensate and the non-condensed fraction of the gas, and only appear in the particle number preserving Bogoliubov approach. The remaining terms in $C^{(2)}$ describe correlations inside the non-condensed cloud.

%being generated by particle number conservation, leading to the depletion of the single mode part of the condensate \cite{footnote7}. 

\subsection{Numerical solution}
\label{subsec:numerics}

To evaluate   the expectation value \eqref{nk} and the correlation functions \eqref{corrterm1} and \eqref{corrterm2}, we first need to compute $\varphi_0$ by solving the inhomogeneous Gross-Pitaevskii
 equations  \eqref{GP} numerically,  and  we then have to determine the  spectrum of $\mathcal{L}_{GP}$. For this purpose, 
 we shall expand all wave functions in terms of two dimensional harmonic oscillator eigenfunctions~\cite{numerics}. 

As a first step, we introduce the dimensionless variables~\cite{Shlyapnikov} 
$$
\zeta=\dfrac{\hbar\omega}{2\mu},\quad y_i=\dfrac{x_i}{R_c},
$$
with $R_c=\sqrt{{2\mu}/{m\omega^2}}$ denoting the  size of the condensate, and rewrite all equations in terms of dimensionless parameters.
 The dimensionless condensate wave function $\phi_0$ of   $N$ bosons can then be expressed as 
\begin{equation*}
\phi_0(\mathbf{y})\equiv\sqrt{N}\,R_c\,\varphi_0(\mathbf{y}\, R_c).
\end{equation*}
This function is normalized to $N$ and, by Eq.~\eqref{GP}, minimizes the dimensionless energy functional
%units /(\frac{\hbar^2 }{2 m R_C^2}) 
\begin{eqnarray}
\mathcal{E}_0 &=&
\int \!\textmd{d}^2\mathbf{y}\! \Bigl (\zeta^2|\nabla_{\mathbf{y}}\phi_0(\mathbf{y})|^2 + (\mathbf{y}^2-1)
|\phi_0(\mathbf{y})|^2
\nonumber 
\\
&& \phantom{n}+\dfrac{g}{2\mu R_c^2}\,|\phi_0(\mathbf{y})|^4\Bigr).
\nonumber 
\end{eqnarray}
We can therefore determine it by expanding $\phi_0(\mathbf{y})$ in terms of  $d=2$ dimensional  harmonic oscillator eigenfunctions,
\begin{equation*}
\phi_0(y)=\sum_{k=0}^{k_{\rm cut}} a_k e^{-\frac{y^2}{2\zeta}}L_k\left(\frac{y^2}{\zeta}\right),
\end{equation*}
with $L_k$  the $k$'th Laguerre-polynomial and $k_{\rm cut}$ finite cutoff introduced for numerical calculations, and then by determining the coefficients $a_k$  via the gradient method. 

Having the  condensate wave function $\phi_0$ at hand, we determine 
 the Bogoliubov eigenfunctions $u_s(\mathbf{x})$ and $v_s(\mathbf{x})$ by solving the   eigenvalue equation of $\mathcal{L}_{GP}$. 
 In order to take into account the projection $\hat{Q}_0$ in Eq. \eqref{LGP}, we modify $\mathcal{L}_{GP}$ 
by a 'Lagrange multiplier' 
\begin{align}
&\mathcal{L}^\prime_{GP}=\begin{pmatrix}
\mathcal{H}+ g\,N\, |\varphi_0|^2+\lambda P_0 & g\,N\,\varphi_0^2 \\
-g\,N \,(\varphi_0^*)^2 & \!\!\! -\mathcal{H}- g\,N\,|\varphi_0|^2+\lambda P_0
\end{pmatrix},
\label{LGP_mod}
\end{align}
with $\hat{P}_0\equiv|\varphi_0\rangle\langle\varphi_0|$ denoting the projection to the condensate wave function and  $\mathcal{H}$
 the mean field Hamiltonian,  given by Eq. \eqref{meanH}. The 
parameter $\lambda$ is  chosen to be large enough to ensure that the low energy eigenfunctions of $\mathcal{L}^\prime_{GP}$, 
 orthogonal to $\varphi_0$, be clearly separated from the high energy spectrum, having finite overlap with the condensate wave function. By keeping only the eigenfunctions of low eigenvalues, annihilated by $\hat{P}_0$, we can determine the excitation spectrum and eigenvectors of the original projected 
 Bogoliubov operator $\mathcal{L}_{GP}$.

Similar to $\phi_0$, we  determine the  eigenfunctions $u_s(\mathbf{x})$ and $v_s(\mathbf{x})$  from the eigenvalue equation of $\mathcal{L}^\prime_{GP}$
by expanding  them in terms of oscillator eigenfunctions. The calculation can be simplified by making use of the rotational symmetry of the condensate, and treating  sectors with different angular momenta $m$ separately. Eigenvectors can then be  classified using radial and angular momentum  indices, $s=(n,m)$, and the eigenfunctions can be expanded in polar coordinates  as 
\begin{align}\label{uv}
&\begin{pmatrix}
u_{n,m}(\mathbf{y}) \\ v_{n,m}(\mathbf{y})
\end{pmatrix}= \nonumber\\
&\quad\quad\sum_{k=0}^{k_{\rm cut}}
\begin{pmatrix}
\alpha^{(m)}_{n k} \\ \beta^{(m)}_{n k}
\end{pmatrix}
e^{i m\varphi}\left(\dfrac{y}{\sqrt{\zeta}}\right)^{|m|}L_{k}^{|m|}\left(\frac{y^2}{\zeta}\right) e^{-\frac{y^2}{2\zeta}}\;,
\end{align}
with $L_{k}^{|m|}$ denoting the generalized Laguerre polynomial of indices $k$ and $|m|$. Substituting this expression into the eigenvalue equations \eqref{LGP_mod}  allows us to determine the coefficients $\alpha^{(m)}_{n k}$ and $\beta^{(m)}_{n k}$. 
Finally, as a last step, we  can now take the Fourier transform of the functions $\phi_0(\mathbf{y})$, $u_s(\mathbf{y})$ and $v_s(\mathbf{y})$ numerically and
 evaluate the expectation value $\langle \hat n_{\mathbf{k}}\rangle$ and the correlation function $C(\mathbf{k},\mathbf{k}^{\prime})$~\cite{footnote3}.

\begin{figure}[t]
\includegraphics[trim=1 0 0 0,clip=true,width=0.9\columnwidth]{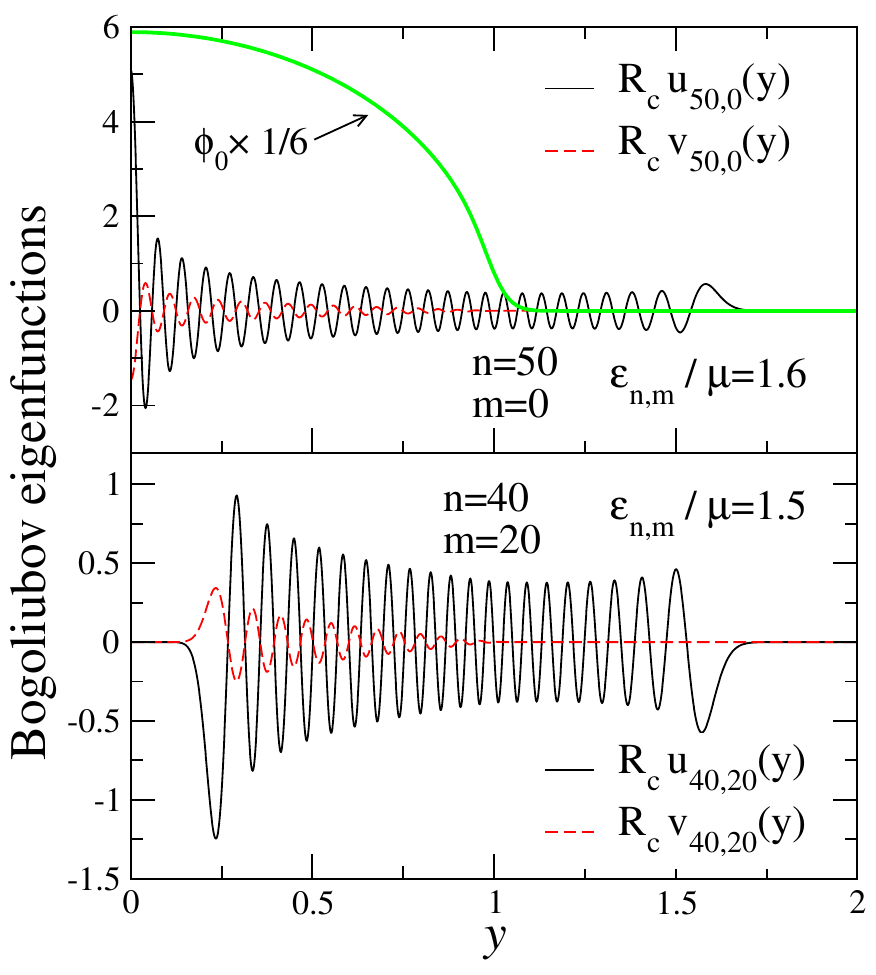}
\caption{Radial part of  the  dimensionless Bogoliubov eigenfunctions $R_cu_{n,m}(\mathbf{x})$, $R_cv_{n,m}(\mathbf{x})$ plotted 
as a function of the dimensionless radial coordinate $y=|\mathbf{x}| R_c$  for $(n,m)=(50,0)$ (top) and $(n,m)=(40,20)$ (bottom), corresponding to excitation energies $\varepsilon_{50,0}/\mu=1.6$ and $\varepsilon_{40,20}/\mu=1.5$ respectively. 
Here $R_c=\sqrt{2\mu / (m \,\omega^2)}$ is the typical size of the condensate, $\zeta^{-1}= {2\mu}/(\hbar\omega)=100$ and $\mu R_c^2/g=1250$, corresponding to  $N=1962$ particles and $\langle \delta \hat{N}\rangle =608$. In the top figure, the dimensionless single-mode condensate wave function $\phi_0$ is also displayed.   The anomalous part $v_{n,m}$ is nonzero only in the regime of the condensate, while the normal part $u_{n,m}$ 
of the wave function can be more extended.  For $m\neq 0$ both $u_{n,m}\rightarrow 0$ and $v_{n,m}\rightarrow 0$ at the center of the trap.
}
\label{fig:uvx}
\end{figure}

\begin{figure}[t]
\includegraphics[trim=1.5 0 0 0,clip=true,width=0.9\columnwidth]{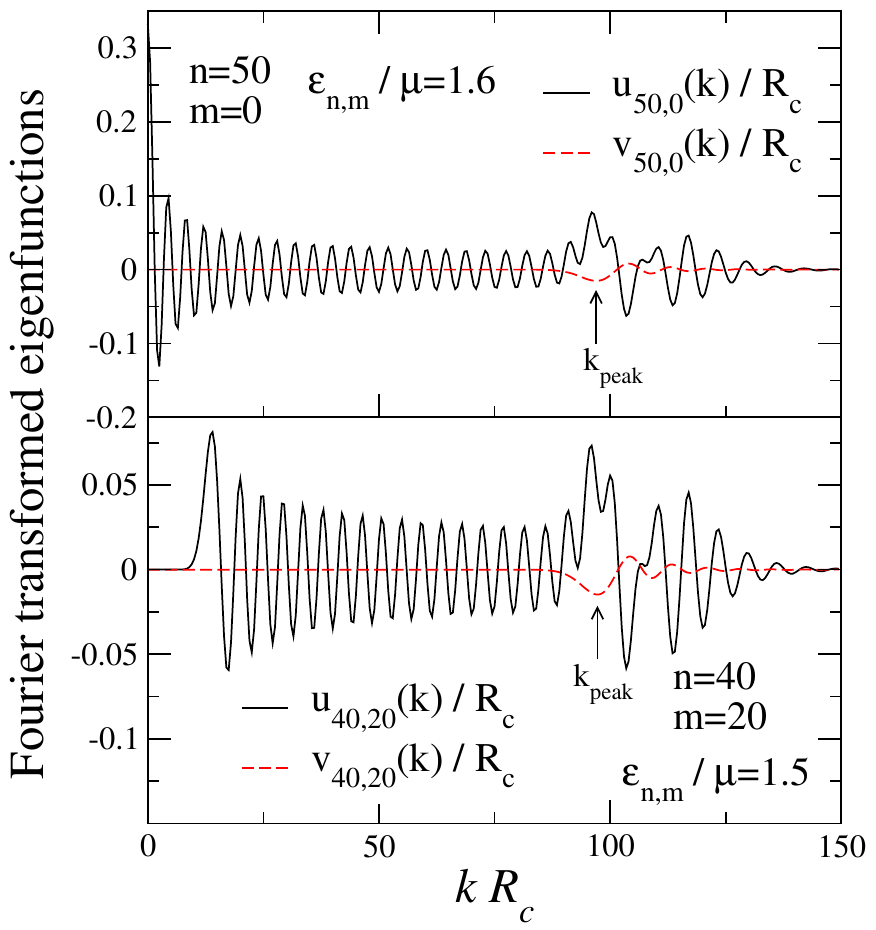}
\caption{Radial part of the dimensionless Fourier transformed Bogoliubov eigenfunctions $u_{n,m}(\mathbf{k})/R_c$, $v_{n,m}(\mathbf{k})/R_c$  as a function of the dimensionless wave number $|\mathbf{k}| R_c$ for $(n,m)=(50,0)$ and $(n,m)=(40,20)$, corresponding to excitation energies $\varepsilon_{50,0}/\mu=1.6$ and $\varepsilon_{40,20}/\mu=1.5$ respectively. Here $R_c=\sqrt{2\mu / (m\,\omega^2)}$ typical size of the condensate, $\zeta^{-1}= {2\mu}/(\hbar\omega)=100$, and $\mu R_c^2/g=1250$, corresponding to $N=1962$ particles and $\langle \delta \hat{N}\rangle =608$. The anomalous component $v_{n,m}(\mathbf{k})$ has a well defined peak at wave number $|\mathbf{k}_{\textmd{peak}}|$ and vanishes for lower $|\mathbf{k}|$, while the normal part $u_{n,m}(\mathbf{k})$ is extended in momentum space.}
\label{fig:uvk}
\end{figure}

\section{Results}
\label{sec:result}

\subsection{Wave functions}

Typical examples of  the condensate wave functions and the radial parts of the Bogoliubov eigenfunctions 
 are shown  in Fig.~\ref{fig:uvx}. The anomalous component of the quasiparticle wave function, $v_{n,m}(y)$, originates from the interaction with the single-mode part 
 of the condensate, and its support  is determined by the extension of the latter. In contrast, the normal 
component $u_{n,m}(y)$ is not constrained to the regime  $\varphi_0\ne 0$, and for high energy quasiparticles it resembles to a harmonic oscillator wave function. Furthermore, as the corresponding excitation energy $\varepsilon_{n,m}$ increases, the interaction energy becomes negligible compared to the kinetic and potential energies, leading to a decrease in the amplitude of $v_{n,m}(y)$.

The Fourier transforms of the radial parts of the  eigenfunctions are plotted as a function of the dimensionless 
wave number $|\mathbf{k}|\,R_c$ in Fig.~\ref{fig:uvk}. The normal component $u_{n,m}(k)$ involves many momenta, and is
therefore  quite extended in  Fourier space. The Fourier transform  of the anomalous component $v_{n,m}(k)$,
however, exhibits a well-defined peak at  $\mathbf{k}_{\rm peak}$. This is explained by the fact that  
$v_{n,m}(y)$ is constrained to the regime where the condensate is present,  
and  there it  oscillates  with an approximately constant radial wave number,  $\mathbf{k}\approx \mathbf{k}_{\rm peak}$.

\subsection{Particle number distributions}

\begin{figure}[t]
\includegraphics[width=\columnwidth]{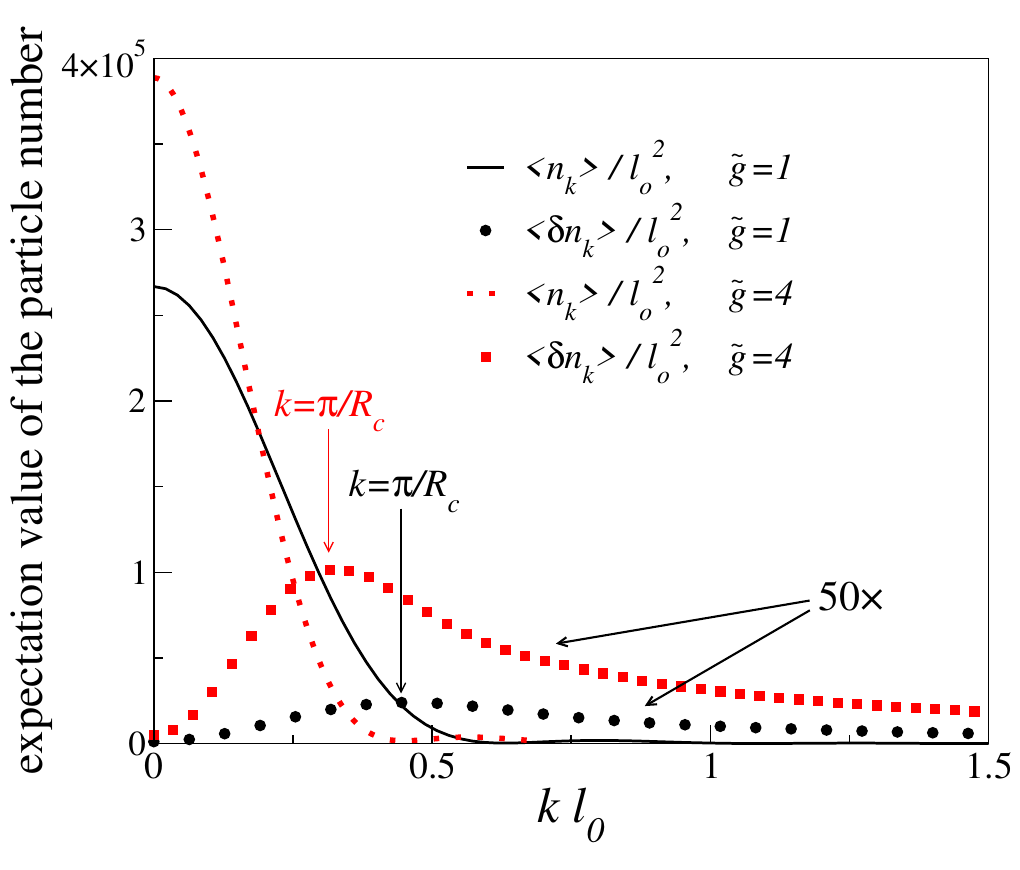}
\caption{Dimensionless expectation values $\langle\hat n_{\mathbf{k}}\rangle / l_0^2$  as a function of $|\mathbf{k}|\,l_0$ for  $N=1962$ and for dimensionless interaction strengths $\tilde{g}=1$ and $\tilde{g}=4$, corresponding to $\langle\delta\hat{N}\rangle=145$ and $\langle\delta \hat{N}\rangle=608$. Dotted lines represent 
contributions of non-condensed particles $\langle\delta\hat n_{\mathbf{k}}\rangle / l_0^2$, with $l_0=\sqrt{\hbar/(m\omega)}$, multiplied by a factor of 50 for better visibility. The extension of the condensate increases with increasing $\tilde{g}$, and the peak in $\langle\hat n_{\mathbf{k}}\rangle $ gets narrower. The long tail quasiparticle contributions 
$\langle\delta\hat n_{\mathbf{k}}\rangle$ get more pronounced with increasing $\tilde{g}$.
}
\label{fig:nk}
\end{figure}

\begin{figure}[t]
\includegraphics[width=\columnwidth]{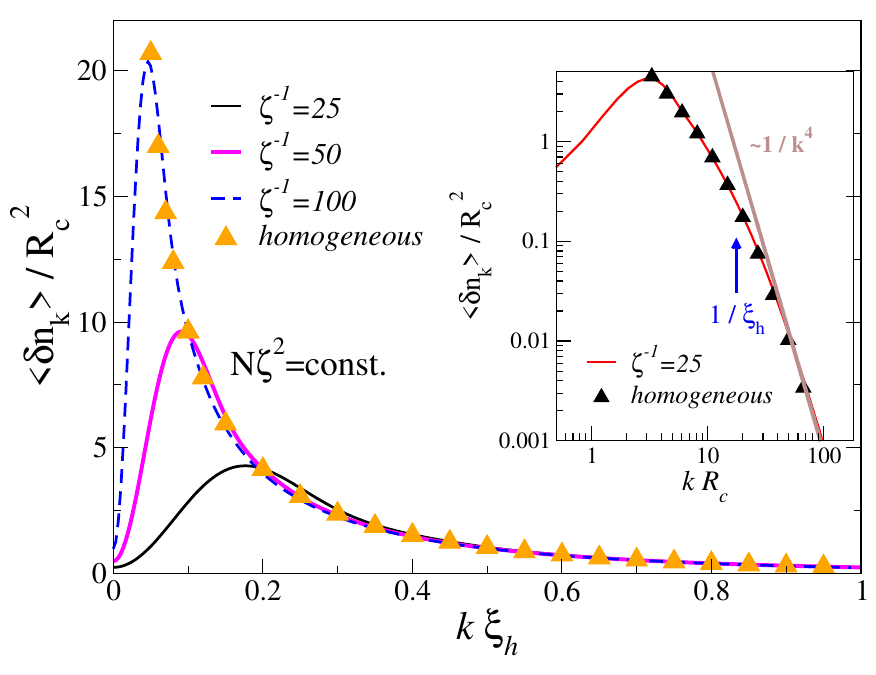}
\caption{Scaling collapse of  $\langle\delta\hat{n}_{\mathbf{k}}\rangle/R_c^2$, plotted as a function of $\mathbf{k}\,\xi_h$ for different $\zeta=\hbar\omega/(2\,\mu)$'s, while keeping $\tilde{g}=4$  and $\rho(0)$ constant. Here $R_c=\sqrt{2\mu / (m \,\omega^2)}$ is the typical size of the condensate, $\xi_h=\hbar/\sqrt{m\,\mu}$ is the healing length with $\mu=g\rho(0)$, and we used $\zeta^{-1}=25$, $\zeta^{-1}=50$ and $\zeta^{-1}=100$, corresponding to $(N,\,\langle\delta\hat{N}\rangle)=(121,34)$, $(N,\,\langle\delta\hat{N}\rangle)=(489,145)$ and $(N,\,\langle\delta\hat{N}\rangle)=(1962,608)$ respectively. The homogeneous momentum distribution, Eq. \eqref{eq:nkhom}, is also plotted for comparison, yielding good agreement with the common envelope function traced out by $\langle\delta\hat{n}_{\mathbf{k}}\rangle/R_c^2$ as $\omega$ decreases. Inset: non-condensed contribution $\langle\delta\hat{n}_{\mathbf{k}}\rangle/R_c^2$, plotted as a function of $\mathbf{k}\,R_c$ for $\tilde{g}=4$ and $\zeta^{-1}=25$, using logarithmic scale on both axis. Homogeneous distribution, Eq. \eqref{eq:nkhom}, is also shown. For large wave numbers $|\mathbf{k}|\gg 1/\xi_h$, the universal power law decay $\sim 1/|\mathbf{k}|^4$ is recovered.
}
\label{fig:dnk}
\end{figure}

 The expectation values of the particle number $\hat{n}_\mathbf{k}$, determined from Eq.\eqref{nk}, are plotted in Fig. \ref{fig:nk} for different dimensionless interaction strengths $\tilde{g}$.   The contribution $\langle\delta\hat{n}_\mathbf{k}\rangle$ of the non-condensed particles  is  shown separately. 
 The  expectation values are dominated by   the single mode part of condensate, giving rise to a large and narrow peak at small wave numbers, $|\bk|\lesssim 1/R_c$. 
 Increasing $\tilde g$ amounts in more extended condensate wave functions in real space, and thereby a  narrower peak in 
 $\langle\hat{n}_\mathbf{k}\rangle$. The non-condensed fraction, $\langle\delta\hat{n}_\mathbf{k}\rangle$, gives only a negligible correction for
small momenta,    $|\mathbf{k}|\lesssim 1/R_c$. 
However, it decays approximately as $1/|\mathbf{k}|$, much more slowly than the central condensate peak, and 
 dominates the \emph{large} wave number behavior, $|\mathbf{k}|> 1/R_c$. For even larger values beyond the 
 inverse healing length,  $|\mathbf{k}|\gg\sqrt{m\mu}/\hbar\equiv\xi_h^{-1}$, 
  $\langle\delta\hat{n}_{\mathbf{k}}\rangle$  goes rapidly 
to zero in a universal fashion as $\sim 1/|\mathbf{k}|^4$ ~\cite{viverit,tan,1/k4} (see also Fig.~\ref{fig:dnk}). Although small in amplitude, %this tail
the contribution from $\delta n_{\mathbf{k}}$
 hosts about $\sim\;30\%$ of the particles for the interactions considered here. Increasing $\tilde g$  
 depletes the condensate  further and leads to a gradual  increase in $\langle\delta\hat{n}_\mathbf{k}\rangle$.

The expectation value of the non-condensed fraction, $\langle\delta\hat{n}_\mathbf{k}\rangle$, is investigated in more detail in Fig. \ref{fig:dnk}, where we compare our numerical results with the momentum distribution of a homogeneous gas. Decreasing the  trapping frequency $\omega$, while keeping the density of the condensate at the center of the trap and the interaction strength (or, equivalently, the healing length $\xi_h=\hbar/\sqrt{m\mu}$)   constant, 
amounts in  a slowly varying  condensate wave function in a wide central region. Therefore, in this limit, a homogeneous system is expected to yield a good approximation for the non-condensed fraction $\langle\delta\hat n_{\mathbf{k}}\rangle$. To make a precise  comparison, however, we need to keep in mind that 
$ n_{\mathbf{k}}$ is dimensionful, and scales as $ n_{\mathbf{k}} \sim (\text{length})^2$. In our case, the size of the condensate $R_c$ 
plays the role of the system size $L$ of a homogeneous system. Therefore,  
to recover  the homogeneous result, we need to investigate the dimensionless expectation value $\langle\delta\hat{n}_{\mathbf{k}}\rangle/R_c^2$. Since the density of the condensate at the center of the trap %, $\rho(0)$, 
scales as $\rho(0)\sim N/R_c^2\sim N\zeta^2/\xi_h^2$, we calculated $\langle\delta\hat{n}_{\mathbf{k}}\rangle/R_c^2$ for different $\zeta$ values, while keeping $N\zeta^2$ and $\xi_h$ constant.  
As shown in Fig. \ref{fig:dnk}, with decreasing $\omega$, the height of the peak in $\langle\delta\hat{n}_{\mathbf{k}}\rangle/R_c^2$ scales as $\sim 1/\omega$, and the peak position shifts to smaller wave numbers, such that  the high momentum part traces out a common envelope function, 
%G 
 just the momentum distribution of a homogeneous gas.

The momentum distribution of a homogeneous system of size $R_c$ and density $\rho_0$ is given by \cite{CastinReview}
\begin{equation}\label{eq:nkhom}
\dfrac{\langle\delta\hat n_{\mathbf{k}}\rangle_{\mathrm{hom}}}{R_c^2\, \pi}=\dfrac{1}{2}\left(\dfrac{(k\xi_h^0)^2+2}{\sqrt{(k\xi_h^0)^2((k\xi_h^0)^2+4)}}-1\right),
\end{equation}
with $\xi_h^0=\hbar/\sqrt{mg\rho_0}$  the healing length of the homogeneous gas, and $R_c^2\,\pi$  the volume of the cylindrically symmetric system. To make a quantitative comparison with our numerical results, plotted in Fig. ~\ref{fig:dnk}, to Eq. \eqref{eq:nkhom}, we have chosen $\rho_0$ as the average density of the inhomogeneous trapped gas. In the limit of small confining frequency $\omega$, the  condensate is well described by the Thomas-Fermi profile, yielding $\rho_0 = \rho(0)/2$.
%\begin{equation*}
%\rho_0\equiv\rho_{\rm average}=\dfrac{\rho(0)}{R_c^2\,\pi}\int_0^{R_c }{\rm d}r\, 2\pi r\,(1-r^2/R_c^2)=\rho(0)/2,
%\end{equation*} 
%with $\rho(0)=\mu/g$ denoting the density at the center of the trap.

%We plotted the homogeneous result, Eq.\eqref{eq:nkhom}, with $\rho_0=\rho(0)/2$ in Fig.~\ref{fig:dnk}. 

We find good agreement with the common envelope function without any further  fitting parameter. 
%, we found good agreement with the common envelope function traced out by the inhomogeneous density distribution, 
% $\langle\delta\hat{n}_{\mathbf{k}}\rangle/R_c^2$, as the trapping frequency is decreased. As mentioned earlier, and also confirmed by Eq. \eqref{eq:nkhom}, 
The non-condensed contribution, $\langle\delta\hat{n}_{\mathbf{k}}\rangle$, decays as $\sim 1/|\mathbf{k}|$ for wave numbers $1/R_c\ll |\mathbf{k}|\ll 1/\xi_h$, while for even larger momenta, $|\mathbf{k}|\gg 1/\xi_h$, the expected $\sim 1/|\mathbf{k}|^4$ decay is recovered (see inset of Fig. \ref{fig:dnk})~\cite{viverit,tan,1/k4}. 
%The latter regime, $|\mathbf{k}|\gg 1/\xi_h$, is plotted separately in the inset of Fig. \ref{fig:dnk}, using logarithmic scale on 
%both axis, confirming the power law behavior. The homogeneous distribution Eq. \eqref{eq:nkhom} is also shown, again giving well 
%agreement with numerical data. 

\subsection{Correlation functions}
\label{sub:correlation_functions}

\begin{figure*}[t!]
\includegraphics[width=0.75\textwidth]{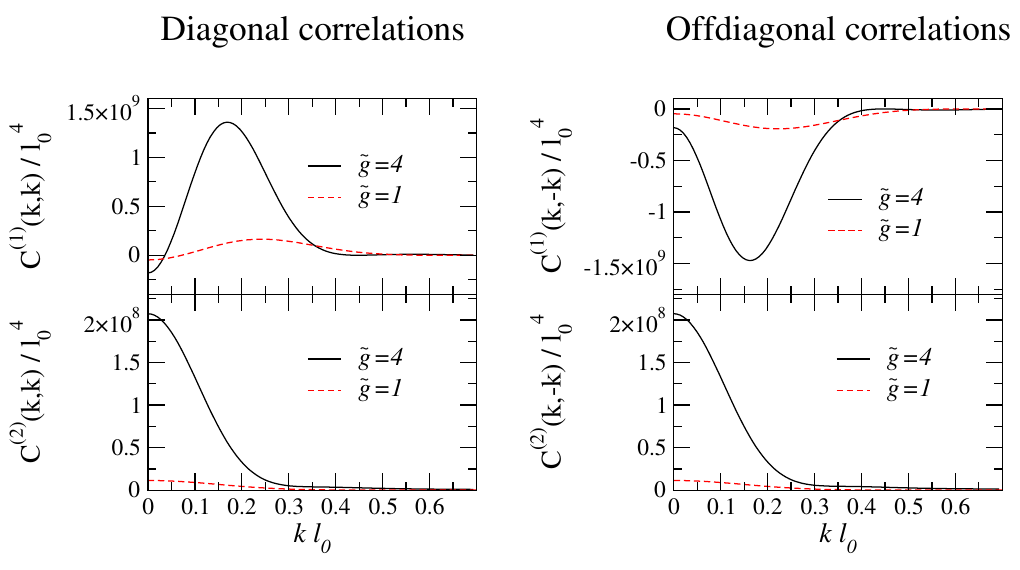}
\caption{ Different contributions to dimensionless diagonal and offdiagonal correlation functions $C(\mathbf{k},\mathbf{k})/l_0^4$ and $C(\mathbf{k},-\mathbf{k})/l_0^4$, plotted as a function of dimensionless wave number $|\mathbf{k}|\,l_0$ for fixed $N=1962$ and for two different interaction strength $\tilde{g}=1$ and $\tilde{g}=4$. Here $l_0=\sqrt{\hbar/(m\,\omega)}$ is the oscillator length, and the interaction values correspond to $\langle\delta\hat{N}\rangle=138$ and $\langle\delta \hat{N}\rangle=608$ respectively. The condensate-quasiparticle contribution $C^{(1)}$ gives a positive peak in diagonal correlations, but gets negative in the offdiagonal, expressing that quantum fluctuations deplete the condensate. As in a homogeneous system, the quasiparticle-quasiparticle correlation $C^{(2)}$ is positive both in the diagonal and in the offdiagonal. However, this contribution is much smaller than $C^{(1)}$ for wave numbers of the order of $1/R_c$. The amplitude of the correlations $C^{(1)}$ and $C^{(2)}$ increases with increasing interaction strength, as the hybridization of the condensate with virtual excitations gets more pronounced.}\label{fig:corr}
\end{figure*}

In Section~\ref{subsec:bogoliubov}, we derived  the correlation function $C(\bk,\bk')=\langle\delta\hat n_\bk \delta\hat n_{\bk'} \rangle $ 
within the particle number conserving Bogoliubov approach, and separated the leading ($\sim |\delta\psi|^2$) and 
subleading  ($\sim |\delta\psi|^4$) contributions from the leading shot noise signal in the terms $C^{(1)}(\bk,\bk')$
and $C^{(2)}(\bk,\bk')$, respectively.
%
%Similar to $\langle\hat n_{\bk}\rangle$, it is instructive to divide the correlation function $C(\bk,\bk')=\langle\delta\hat n_\bk \delta\hat n_{\bk'} \rangle $  into %two parts. As in Eq.~\eqref{corrterm1}, we separate the leading order contributions proportional to the total particle number, expressing cross-correlations %between the quantum fluctuation induced quasiparticles and the condensate. The remaining correlation terms, contained in Eq.~\eqref{corrterm2}, involve %correlations inside the non-condensed cloud, and corrections to the condensate-quasiparticle correlations, generated by particle number conservation. 
%
These contributions, given by Eqs.~\eqref{corrterm1} and ~\eqref{corrterm2}, are plotted in Fig. \ref{fig:corr}  for wave numbers $\mathbf{k^\prime}=\mathbf{k}$ and $\mathbf{k^\prime}=-\mathbf{k}$ for various interaction strengths $\tilde{g}$. The variance of the particle number $\hat{n}(\mathbf{k})$ is given by the sum of the singular shot noise term and the diagonal correlations $C(\bk,\bk)$, so the diagonal part $C(\bk,\bk)$ is not necessarily positive. However, the off-diagonal part 
 $C(\bk,-\bk)$ develops a more pronounced anticorrelation dip, due to the depletion of the condensate by quasiparticle excitations.

\begin{figure}[t]
\includegraphics[width=\columnwidth]{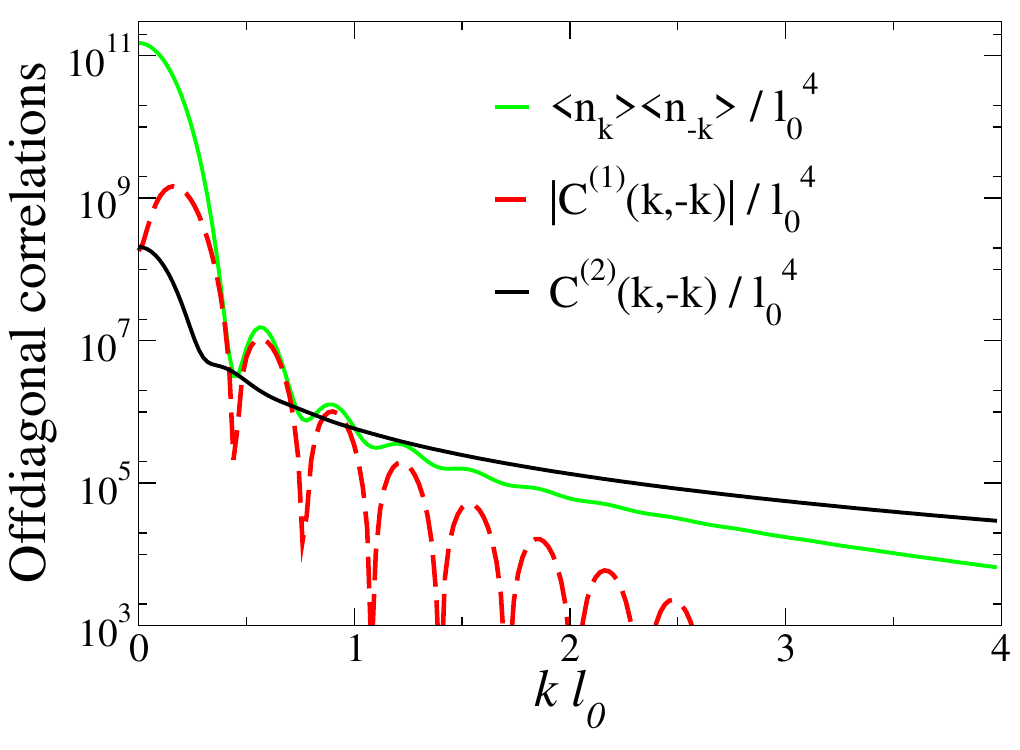}
\caption{Different contributions to dimensionless offdiagonal correlation function $C(\mathbf{k},-\mathbf{k})/l_0^4$, plotted as a function of dimensionless wave number $|\mathbf{k}|\,l_0$ for particle number $N=1962$ and interaction strength $\tilde{g}=4$, using logarithmic scale on vertical axis. Here $l_0=\sqrt{\hbar/(m\,\omega)}$ is the oscillator length, and the interaction corresponds to $\langle\delta \hat{N}\rangle=608$. The background signal $\langle\hat n_{\mathbf{k}}\rangle\langle\hat n_{-\mathbf{k}}\rangle/l_0^4$ shows a steep decrease due to the disappearance of condensate wave function, followed by a slower decay as an effect of non-condensed particles. The condensate-quasiparticle contribution $C^{(1)}$ is constrained to the regime of the single-mode condensate, and converges to zero rapidly for $|\mathbf{k}|\gg 1/R_c$. The quasiparticle-quasiparticle correlation $C^{(2)}$ gives a slowly decaying tail, dominating the correlation function for $|\mathbf{k}|\gg 1/R_c$.}
\label{fig:corrlog}
\end{figure}

The non-connected part $\langle \hat{n}_\mathbf{k}\rangle\langle\hat{n}_\mathbf{k^\prime}\rangle$ of the correlator $\langle \hat{n}_\mathbf{k}\hat{n}_\mathbf{k^\prime}\rangle$
does not distinguish between diagonal and offdiagonal correlations, and follows readily from Fig. \ref{fig:nk}. 
 Although this large signal is subtracted in the correlation function,   Eq.~\eqref{corr}, it still provides a large background in an experiment  
and  may therefore be hard to separate it from the more interesting part of the signal (see Fig.~\ref{fig:corrlog}). 
Similar to $\langle \hat{n}_\mathbf{k}\rangle$, the product $\langle \hat{n}_\mathbf{k}\rangle\langle\hat{n}_\mathbf{k^\prime}\rangle$  exhibits a sharp peak with typical width 
$|\mathbf{k}'| \sim |\mathbf{k}|\sim 1/R_c$, originating from the single-mode condensate, also shown in Fig. \ref{fig:nk}. 
The expectation values $\langle \hat{n}_\mathbf{k}\rangle$ being invariant under rotations, 
$\langle \hat{n}_\mathbf{k}\rangle\langle\hat{n}_\mathbf{k^\prime}\rangle$  is clearly also independent of the relative directions of $\bk$ and $\bk'$, and is 
'cylindrically' symmetrical.

The  leading contribution $C^{(1)}$, shown in the top panels of Fig. \ref{fig:corr},  
accounts for correlations between the single-mode condensate and the non-condensed fraction of the gas. 
Consequently, similar to $\varphi_0(\bk)$,   $C^{(1)}$ is constrained to small wave numbers, 
and decreases rapidly for $|\mathbf{k}|>1/R_c$.  The function 
$C^{(1)}$ exhibits an \emph{anticorrelation dip} in the off-diagonal $\bk'\approx -\bk$
 for wave numbers $|\bk|\sim 1/R_c$. This dip dominates the small momentum 
  behavior of $C(\bk,\bk')$,  and  gets    more pronounced for increasing interaction strength. 
 The negative correlation observed originates from particle number preserving processes, where the interaction 
  $g$ creates  quasiparticle pairs from the condensate. The coherent transfer of these particle pairs between the single-mode condensate and the non-condensed fraction of gas is responsible for the anticorrelation dip in $C^{(1)}$ (see also Section~\ref{sub:simple}) \cite{footnote6}.
  %G
  Notice that this anticorrelation also appears in the standard grand canonical Bogoliubov approach: there the factors $\varphi_0(\mathbf{k})$ 
  and $\varphi_0(\mathbf{k^\prime})$ in the first four terms of Eq. \eqref{corrterm1} emerge as the coherence factors of the condensate, and $\varphi_0$ and $\varphi_0^*$ correspond to removing or adding a particle to the condensate.
 Therefore, these terms  can be associated with 
  particle number conserving processes, captured to a certain degree already 
by  the usual (non-conserving) Bogoliubov approach.  
 
 %In that case, thermal fluctuations play a somewhat similar role as  interaction induced quantum 
 %fluctuations do in our zero temperature system. 

Finally,  the contribution  $C^{(2)}$, shown in the bottom panels of Fig.~\ref{fig:corr},
 describes correlations within the non-condensed (more precisely, 
non single-mode condensed) cloud, but also incorporates  contributions 
arising within the  particle number conserving Bogoliubov approach, generated by the term 
$-|\varphi_0(\mathbf{k})|^2\,\delta\hat{N}$ in the expression of $n_{\mathbf{k}}$, Eq.~\eqref{n_k}.
 These latter contributions give rise to  a  central peak of  width $\sim 1/R_c$, 
 and yield  a  small correction to the leading order correlations between 
the single-mode condensate and the non-condensed particles, contained in 
$C^{(1)}$. Correlations within the non-condensed fraction, captured by the other terms in $C^{(2)}$, 
result in a \emph{slowly decaying positive correlation tail} both in the
 diagonal, $\mathbf{k^\prime}=\mathbf{k}$, and in the offdiagonal, 
$\mathbf{k^\prime}=-\mathbf{k}$. 
This positive correlation is qualitatively similar to the simple 
Bogoliubov result, valid for weakly interacting homogeneous 
condensates \cite{Altman}. Albeit their contribution is small compared to the amplitude 
of the central peaks in $C^{(1)}$, quantum fluctuations
dominate the correlation function for wave numbers 
$|\mathbf{k}|\gg 1/R_c$, showing that the fluctuating 
part of the ground state consists of \emph{pairs} of quasiparticles, as visualized in Fig. \ref{fig:sketch}. 
The amplitude of this correlation tail is sensitive to interactions, and  is 
further enhanced by increasing interaction strength $\tilde{g}$.

\begin{figure*}[t]
\includegraphics[trim=0 5 0 0,clip,width=0.79\columnwidth]{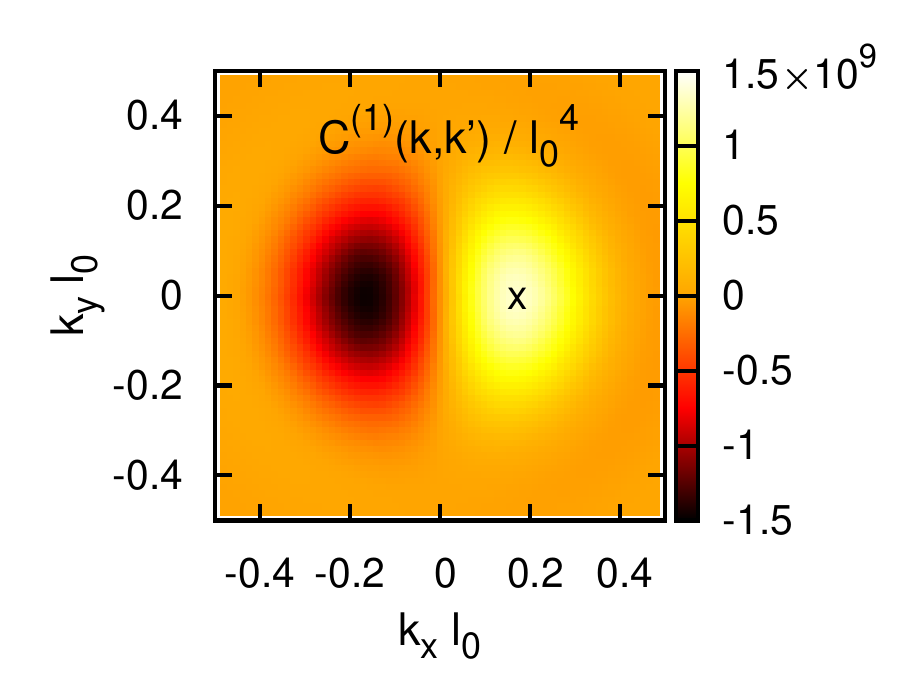}
\includegraphics[trim=0 5 0 0,width=0.79\columnwidth,clip]{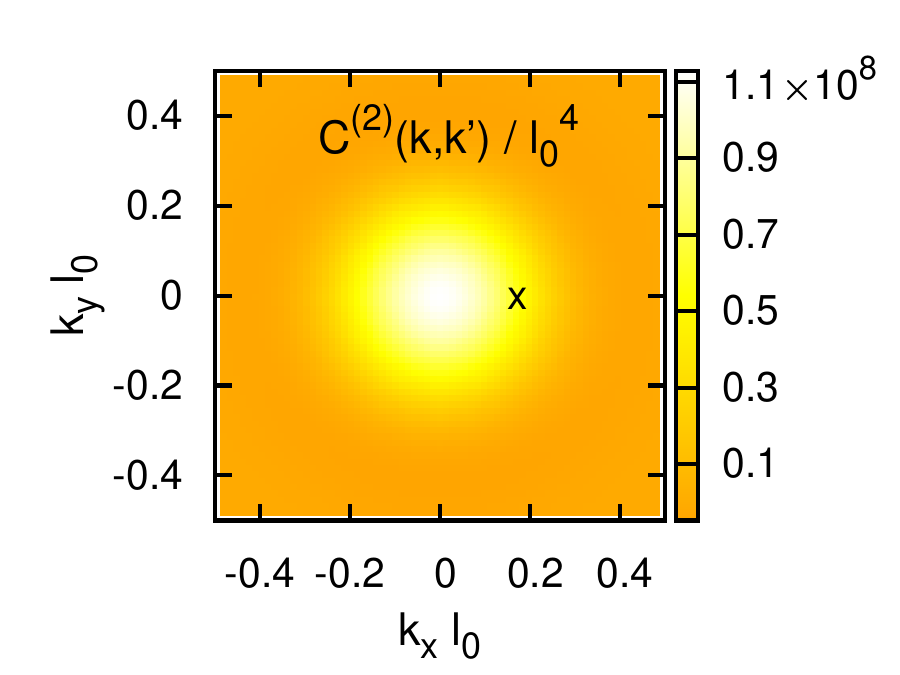} \\
\includegraphics[trim=0 0 5 5,clip,width=0.79\columnwidth]{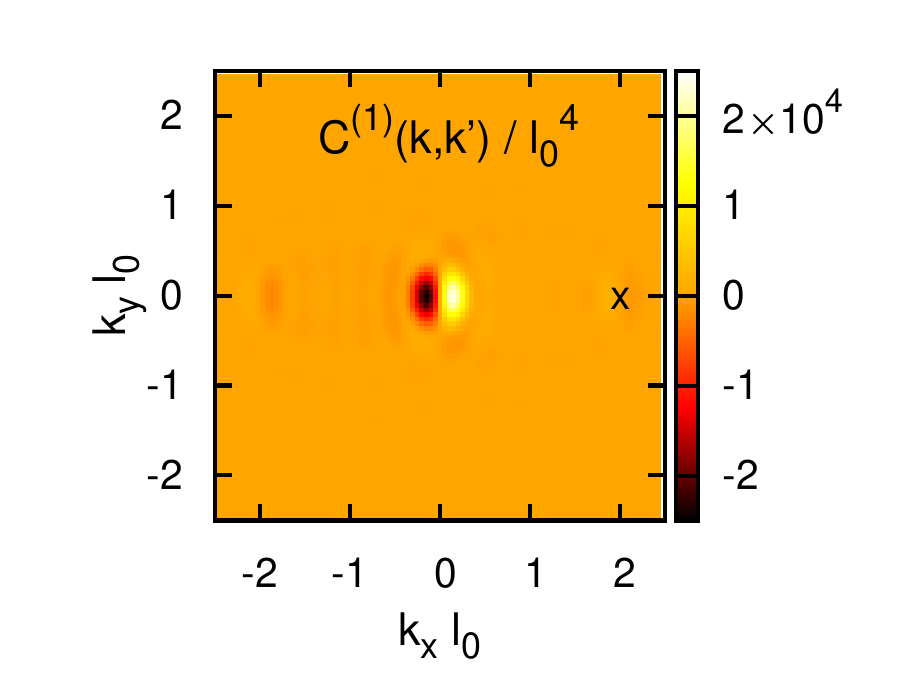}
\includegraphics[trim=0 0 5 5,clip,width=0.79\columnwidth]{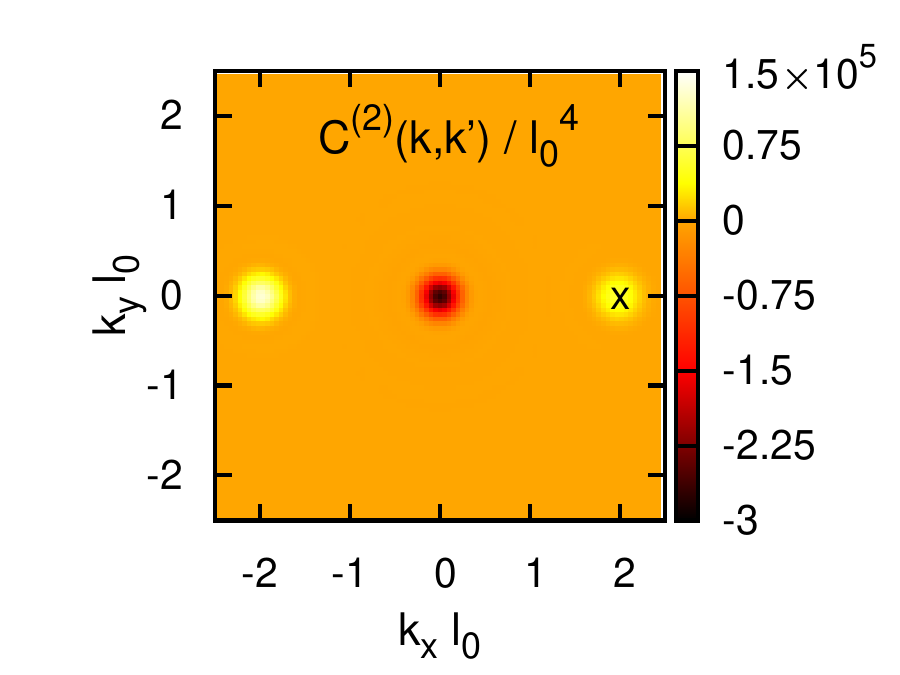}
\caption{
Dimensionless correlation functions $C^{(1)}(\mathbf{k},\mathbf{k^\prime})/l_0^4$ and $C^{(2)}(\mathbf{k},\mathbf{k^\prime})/l_0^4$ plotted as a function of dimensionless wave number $\mathbf{k}\,l_0$, for fixed values of 
$\mathbf{k^{\prime}}$. Here $l_0=\sqrt{\hbar/(m \,\omega)}$ oscillator length, and we have used $\zeta^{-1}=100$ and $\tilde{g}=4$, corresponding to  $N=1962$ particles and $\langle \delta \hat{N}\rangle =608$. First row: $\mathbf{k^{\prime}}\,l_0=(0.16,0)$. The condensate-quasiparticle correlation $C^{(1)}$ is positive if $\mathbf{k}$ and $\mathbf{k^{\prime}}$ point to the same direction, and gives negative contribution in the $\mathbf{k^{\prime}}\approx\mathbf{-k}$ regime. The positive correlation $C^{(2)}$ is concentrated to small $\mathbf{k}\,l_0$ wave numbers, due to subleading corrections to condensate-quasiparticle correlations contained in $C^{(1)}$. Second row: $\mathbf{k^{\prime}}\,l_0=(2,0)$. The dominant contribution here is the quasiparticle-quasiparticle correlation $C^{(2)}$, giving negative values for small wave numbers, and narrow positive peaks around $\mathbf{k}=\mathbf{k^\prime}$ and $\mathbf{k}=-\mathbf{k^\prime}$, expressing correlations in the non-condensed fraction of the gas.}\label{fig:2dcorr}
\end{figure*}

To gain further insight into the structure of $C(\mathbf{k},\mathbf{k^\prime})$,  
we have plotted in
Fig. \ref{fig:2dcorr}
 the correlation functions $C^{(1)}(\mathbf{k},\mathbf{k^\prime})$ 
and $C^{(2)}(\mathbf{k},\mathbf{k^\prime})$,  as  functions of 
$\mathbf{k}$ while keeping $\mathbf{k^\prime}$ fixed. For $|\mathbf{k^\prime}|$ of 
the order of $1/R_c$, opposite to the positive peak at $\mathbf{k}=\mathbf{k^\prime}$, 
an anticorrelation dip arises around the wave number $\mathbf{k}=-\mathbf{k^\prime}$ 
in the condensate-quasiparticle contribution $C^{(1)}$, in accordance 
with the results plotted in Fig. \ref{fig:corr}. 
 This structure, reflecting correlations between the quasiparticles and  the condensate,  
disappears for wave numbers $|\mathbf{k^\prime}|\gg 1/R_c$ (bottom row in Fig. \ref{fig:2dcorr}), where 
 positive correlations  appear for wave numbers $\mathbf{k}$ opposite to $\mathbf{k^\prime}$.

 As shown in the bottom row of Fig. \ref{fig:2dcorr}, for $|\mathbf{k^\prime}|\gg 1/R_c$ two narrow positive peaks 
can be observed in $C^{(2)}$ around wave numbers $\mathbf{k}=\mathbf{k^\prime}$ and $\mathbf{k}=-\mathbf{k^\prime}$.
 These positive contributions originate from pair correlations inside the non-condensed fraction of the gas, and 
are related to the slowly decaying positive tail of the diagonal and off-diagonal correlation function, plotted  
in Fig. \ref{fig:corr}. These pair correlations dominate the tails of ToF images of the condensate. 
 %G
  For small momenta, $|\mathbf{k^\prime}|\sim 1/R_c$, however, the correlation function $C^{(2)}$ 
is dominated by a central peak of typical width $\sim 1/R_c$, originating from subleading, fourth order corrections  in the 
fluctuations $\delta \psi$.
%to the  condensate-quasiparticle correlations contained in $C^{(1)}$, determining  the structure of ToF 
%images at small momenta.

\subsection{Simple model for correlations}
\label{sub:simple}

The structure of the correlation function $C(\mathbf{k},\mathbf{k^\prime})$, discussed above, provides detailed information on the ground state of the system. The  slowly decaying positive tail around $\mathbf{k}=-\mathbf{k^\prime}$ for $|\mathbf{k}|\gg 1/R_c$ is a sign of excitations created in pairs $\mathbf{k}$ and $-\mathbf{k}$, characteristic to the familiar two-mode squeezed structure of the Bogoliubov wave function. On the other hand, the negative off-diagonal correlations found for $|\mathbf{k}|\ll 1/R_c$ show that these pairs of excitations are created coherently from the single mode condensate by quantum fluctuations.

To illustrate the latter point, let us consider the correlations present in two different simple model states, both showing a pair structure of excitations. We first consider a pure state with coherently created excitations, then we calculate the correlations for a mixed state as well, where this coherence is lost. We show that a $p$-wave like structure of the correlation function only emerges in the first case, for coherent quantum fluctuations.

Let us first take the following pure state, with excitations created in pairs,
$$|A\rangle=\left[\left(\hat{b}_0^+\right)^2-g\,\hat{b}_+^\dagger\hat{b}_-^\dagger\right]|0\rangle.$$
Here $\hat{b}_0^\dagger$ denotes a bosonic creation operator, corresponding to the condensate with the cylindrically symmetric wave function $\varphi_0(\mathbf{r})\equiv\varphi_s(r)$. Similarly, $\hat{b}_\pm^\dagger$ represent bosonic fluctuations ($\delta\psi$), orthogonal to $\varphi_0$. By orthogonality they must have a $p$-wave structure in the simplest case: $\varphi_\pm(\mathbf{r})\equiv\varphi_p(r)e^{\pm i \varphi}$, with $(r,\varphi)$ denoting polar coordinates. Indeed, we verified numerically that the excitations with $p$-wave structure, $s=(n,m=\pm 1)$, give rise to the dominant contribution to $C^{(1)}$.

For a small admixture of the $\varphi_\pm$ states, $g\ll 1$, the state $|A\rangle$ can be used as a simple model capturing the $\pm\bf k$ pair structure of the Bogoliubov ground state, with fixed particle number 2. Let us now calculate the correlations induced by $|A\rangle$, $C_A(\mathbf{k},\mathbf{k^\prime})=\langle A|\hat{\psi}^\dagger(\mathbf{k})\hat{\psi}^\dagger(\mathbf{k^\prime})\hat{\psi}(\mathbf{k})\hat{\psi}(\mathbf{k^\prime})|A\rangle$, and inspect the different contributions ordered according to the power of $g$.

Using cylindrical coordinates $\mathbf{k}\leftrightarrow(k,\theta)$, we can express the Fourier transforms of the wave functions $\varphi_{s,\pm}$ as
\begin{align*}
&\varphi_s(\mathbf{k})\equiv\varphi_s(k)=2\pi\int{\rm d}r\, r\,\varphi_s(r)J_0(kr),\\
&\varphi_\pm(\mathbf{k})\equiv -i\,\varphi_p(k)e^{\pm i\theta}=-i\, 2\pi\int{\rm d}r\, r\,\varphi_p(r)J_1(kr)e^{\pm i\theta},
\end{align*}
with $J_0$ and $J_1$ denoting Bessel functions. By using these relations, it is easy to see that the $\sim g^0$ contribution to $C_A(\mathbf{k},\mathbf{k^\prime})$ will be cylindrically symmetric. However, the terms proportional to $g$ will give a contribution
\begin{equation}\label{modelCorr1}
\sim g\,\varphi_s(k)\varphi_s(k^\prime)\varphi_p(k)\varphi_p(k^\prime)\cos(\theta-\theta^\prime).
\end{equation}
This term has the same $p$-wave symmetry, as the condensate-quasiparticle correlation function $C^{(1)}$, and corresponds to positive correlations for $\mathbf{k}=\mathbf{k^\prime}$, but results in an anticorrelation dip for $\mathbf{k}=-\mathbf{k^\prime}$.

The terms proportional to $g^2$ can be divided into a cylindrically symmetric contribution, and an additional term
\begin{equation}\label{modelCorr2}
\sim g^2\,\varphi_p(k)^2\varphi_p(k^\prime)^2\cos(2(\theta-\theta^\prime)).
\end{equation}
As expected from the pair structure built into $|A\rangle$, the $d$-wave symmetry of this contribution is consistent with the large wave number behavior of the quasiparticle-quasiparticle correlation function $C^{(2)}$, resulting in positive correlation for $\mathbf{k}=\pm\mathbf{k^\prime}$. 
%G
At the tails of the ToF image, however,
%$|\mathbf{k}|\gg 1/R_c$, it is no longer true that excitations with $p$-wave structure give the dominant contribution to $C^{(2)}$, instead 
all higher harmonics contribute to the density profile. Repeating the preceding analysis with $\varphi_\pm(\mathbf{r})\equiv\varphi_m(r)e^{\pm i m \varphi}$ for arbitrary $m$ shows that the term proportional to $g^2$ depends on the angles $\theta$ and $\theta^\prime$ as $\cos(2m(\theta-\theta^\prime))$, still leading to positive correlations for % between opposite wave numbers,
 $\theta-\theta^\prime\approx \pi$. 
 To contrast this even structure of $C^{(2)}$%of the quasiparticle-quasiparticle correlation function 
 to the odd \emph{p-wave} symmetry of %the condensate - quasiparticle contribution 
$C^{(1)}$, we  refer to it  as a "$d$-wave"  structure -- in spite of the presence of higher 
 harmonics. 

In order to show, that the contribution given by Eq. \eqref{modelCorr1} can indeed be identified as a sign of coherent quantum fluctuations, let us now consider a mixed state, exhibiting a pair structure similar to $|A\rangle$, described by the density matrix
$$\hat{\rho}=|B\rangle\langle B|+g^2|C\rangle\langle C|,$$
with $|B\rangle=(b_0^\dagger)^2|0\rangle$ and $|C\rangle=b_+^\dagger b_-^\dagger|0\rangle$. The calculation of the correlation function ${\rm Tr}\left(\hat{\rho}\, \hat{\psi}^\dagger(\mathbf{k})\hat{\psi}^\dagger(\mathbf{k^\prime})\hat{\psi}(\mathbf{k})\hat{\psi}(\mathbf{k^\prime})\right)$ shows, that the first order contribution Eq. \eqref{modelCorr1} disappears, while the quasiparticle-quasiparticle term, given by Eq. \eqref{modelCorr2}, persists. Thus the relative phase between the two terms in $|A\rangle$, i.e. the coherence of the interaction induced quasiparticle pairs, is crucial for the anticorrelations observed here and in Ref. \cite{Bouchoule2}.

\section{Conclusion}
\label{sec:cnclusions}

We have studied the momentum distribution and the density correlation function 
of a two-dimensional, harmonically trapped interacting Bose gas. Concentrating on the interplay of quantum fluctuations, confinement and particle number conservation, we performed the calculations at zero temperature, using 
a particle number preserving Bogoliubov-approach.

To characterize the system, we have first calculated the momentum distribution function for various interaction strengths $\tilde{g}$,
which is dominated by a central peak originating from the single-mode condensate.
The amplitude of the non single-mode condensed fraction of the gas is clearly overwhelmed by this central peak. 
However, this latter contribution  is much more extended in Fourier
space, giving a slowly decaying tail. Therefore, it can possibly be disentangled from the single-mode condensate peak experimentally.

By studying the correlation function $C(\mathbf{k},\mathbf{k^\prime})\equiv \langle \delta \hat n_\bk \delta \hat n_{\bk '}\rangle$, we showed that the \emph{anti-correlations} between opposite wave numbers $\mathbf{k}$ and $-\mathbf{k}$, experimentally observed for one-dimensional quasi-condensates \cite{Bouchoule2}, also appear for higher, $d=2$ dimensional systems.
% due to the interplay of particle number conservation and harmonic confinement.
 Moreover, by separating  $C(\mathbf{k},\mathbf{k^\prime})$ into two parts, we identified two distinct contributions to the correlation function, exhibiting different symmetries.

The first contribution, $C^{(1)}$, describing
 correlations between the single-mode condensate and the non-condensed fraction of the gas, is responsible for the development of the anti-correlation dip around $\mathbf{k^\prime}=-\mathbf{k}$. This dip seems to originate from particle number preserving processes, coherently moving particle pairs between the single mode condensate and the non-condensed cloud. For our $d=2$ dimensional system at $T=0$ temperature, the spatial extension of the condensate, $R_c$, takes over the role of thermal wave length $l_\phi$, determining the region of anti-correlations in a one-dimensional quasi-condensate \cite{Bouchoule2}, thus the momentum-space extension of the anti-correlation dip is set by $1/R_c$.
 
In addition to the anticorrelations between nearly opposite wave numbers, $\bk\approx -\bk'$,  mentioned above, $C^{(1)}$  also contains  \emph{forward correlation} for particles of similar momenta,  $\bk\approx \bk'$. The momentum space correlations between the single-mode condensate and the non-condensed fraction of the gas, $C^{(1)}$, thus exhibit  a characteristic \emph{p-wave} structure, and dominate the full correlation function $C(\mathbf{k},\mathbf{k^\prime})$ in the region of small wave numbers $|\bk|,|\bk'|\sim 1/R_c$.

The other part of the correlation function, $C^{(2)}$, stems from correlations within the non-condensed fraction of the gas. It  decays slowly as $\sim 1/|\mathbf{k}|^2$ with a positive tail around the offdiagonal $\mathbf{k^\prime}\approx-\mathbf{k}$, similarly to the Bogoliubov result for homogeneous systems. This contribution exhibits a "$d$-wave"-like symmetry with positive correlations both in the $\bk'\approx\bk$ and $\bk'\approx-\bk$ regimes, and dominates the full correlation function in the region  of large wave numbers, $|\bk|,|\bk'|\gg 1/R_c$, where  short distance correlations at scales $\lambda\ll R_c$ are probed.

%G
The anticorrelations observed seem to rely on several important ingredients: First, they reflect 
the dominant $p$-wave character of the quantum fluctuations, as supported by  a careful analysis of the interaction-induced quantum 
fluctuations~\cite{footnote4}. Second, they evidence  the coherent nature of these quantum fluctuations. 
Finally, they appear to be related to processes, where particles 
 move between the single mode part of the condensate and the fluctuating part, $\delta \psi$.
Indeed,  all important  
features discussed in the previous  paragraphs can be 
captured by a simple  toy model incorporating these three ingredients (see  Section~\ref{sub:simple}). 
The contributions $C^{(1)}$ and $C^{(2)}$ reveal important information about the structure of the interacting superfluid state. The even 
 symmetry of $C^{(2)}$ reflects  that long wave length excitations are created in pairs $\pm\bk$ from the single 
mode condensate, while the \emph{p-wave} structure of $C^{(1)}$ evidences the  
coherence of the quantum  fluctuations.

In actual experiments, one measures the full correlator $\langle \hat{n}_\mathbf{k}\hat{n}_\mathbf{k^\prime}\rangle$ 
instead of the connected part $C(\mathbf{k},\mathbf{k^\prime})$, yielding a large, cylindrically symmetric background signal 
$\langle \hat{n}_\mathbf{k}\rangle\langle\hat{n}_\mathbf{k^\prime}\rangle$. 
This results in a background $\sim N^{1/2}$ times larger than the anti-correlation 
dip in the connected part  around $\pi/R_c$. However, $C^{(1)}$ exhibits a different, \emph{p-wave} symmetry, making its experimental detection possible. 

On the other hand, the positive "$d$-wave"-like tail of $C(\mathbf{k},\mathbf{k^\prime})$ scales as $\sim (N\tilde{g})^2$. Being of the same order of magnitude as the background, it could be
experimentally accessible. To observe these correlations, however, one needs to investigate the tails of the ToF image with momenta $|\bk|\gtrsim 1/R_c$.

\begin{acknowledgments}
This research has been  supported by the National Research, Development and Innovation Office - NKFIH Nos. K105149, SNN118028, K119442 and by the Bolyai Program of the Hungarian Academy of Sciences. ED acknowledges support from Harvard-MIT CUA, NSF Grant No. DMR-1308435, AFOSR Quantum Simulation MURI, AFOSR MURI Photonic Quantum Matter, the Humboldt Foundation, and the Max Planck Institute for Quantum Optics.
\end{acknowledgments}


\begin{references}
\bibitem{footnote0} Low dimensional systems are realized in highly anisotropic traps. When the confinement is removed, the interactions quickly become negligible due to the rapid expansion of the gas in the tightly confined directions.
\bibitem{footnote1} This limit is relevant for  the regime $\rho_{2D} \hbar^2/m k_B T\gg 1$, where phase fluctuations are suppressed.
\bibitem{footnote2} {In reality, the delta potential is not well defined, and a renormalization procedure must be employed. At the mean field level, considered here, however, $g$ can be replaced by its renormalized value.}
\bibitem{footnote5}The last term in $C^{(1)}$, $-N |\varphi_0(\mathbf{k})|^2|\varphi_0(\mathbf{k^\prime})|^2$, originates from the completeness relation $\eqref{complete}$, and ensures that the variance of the total particle number remains zero, $\langle\delta^2\hat{N}\rangle=\int {\rm d}^2\mathbf{k}/(2\pi)^2\,{\rm d}^2\mathbf{k^\prime}/(2\pi)^2\,C(\mathbf{k},\mathbf{k^\prime})=0$. A similar term also appears in the grand canonical description of the condensate, but eventually gets cancelled by the fluctuations in the total particle number, leading to $\langle\delta^2\hat{N}\rangle>0$.
\bibitem{footnote3} {In Eqs. \eqref{nk} and \eqref{corr}  a finite cutoff is introduced in the summation over $s$.}
\bibitem{footnote4} {In one dimension, where these anticorrelations have been experimentally observed, $s$- and $p$-wave fluctuations correspond to  fluctuations in the 'even' and 'odd' channels, respectively.}
\end{references}
\end{document}